\documentclass[trackchanges,twocolumn]{aastex}

\let\tablenum\relax

\usepackage{natbib,epsfig,siunitx,url,graphicx,amsmath,footnote,float,xcolor,cases,lineno}

\begin{document}

\submitted{Accepted for publication in \textit{AJ}}

\title{Dynamical avenues for Mercury's origin I: The lone survivor of a primordial generation of short-period proto-planets}

\author{Matthew S. Clement\altaffilmark{1,2}, John E. Chambers\altaffilmark{1}, \& Alan P. Jackson\altaffilmark{3}}

\altaffiltext{1}{Earth and Planets Laboratory, Carnegie Institution
for Science, 5241 Broad Branch Road, NW, Washington, DC
20015, USA}
\altaffiltext{2}{HL Dodge Department of Physics Astronomy, University of Oklahoma, Norman, OK 73019, USA}
\altaffiltext{3}{School of Earth and Space Exploration, Arizona State University, 550 E Tyler Mall, Tempe, Arizona,85287, USA}

\altaffiltext{*}{corresponding author email: mclement@carnegiescience.edu}

\begin{abstract}

The absence of planets interior to Mercury continues to puzzle terrestrial planet formation models, particularly when contrasted with the relatively high derived occurrence rates of short-period planets around Sun-like stars.  Recent work proposed that the majority of systems hosting hot super-Earths attain their orbital architectures through an epoch of dynamical instability after forming in quasi-stable, tightly packed configurations.  Isotopic evidence seems to suggest that the formation of objects in the super-Earth mass regime is unlikely to have occurred in the solar system as the terrestrial-forming disk is thought to have been significantly mass-deprived starting around 2 Myr after CAI; a consequence of either Jupiter's growth or an intrinsic disk feature.  Nevertheless, terrestrial planet formation models and high-resolution investigations of planetesimal dynamics in the gas disk phase occasionally find that quasi-stable proto-planets with masses comparable to that of Mars emerge in the vicinity of Mercury's modern orbit.  In this paper, we investigate whether it is possible for a primordial configuration of such objects to be cataclysmically destroyed in a manner that leaves Mercury behind as the sole survivor without disturbing the other terrestrial worlds.  We use numerical simulations to show that this scenario is plausible.  In many cases, the surviving Mercury analog experiences a series of erosive impacts; thereby boosting its Fe/Si ratio.  A caveat of our proposed genesis scenario for Mercury is that Venus typically experiences at least one late giant impact.
\end{abstract}

\section{Introduction}

\begin{figure*}
	\centering
	\includegraphics[width=.99\textwidth]{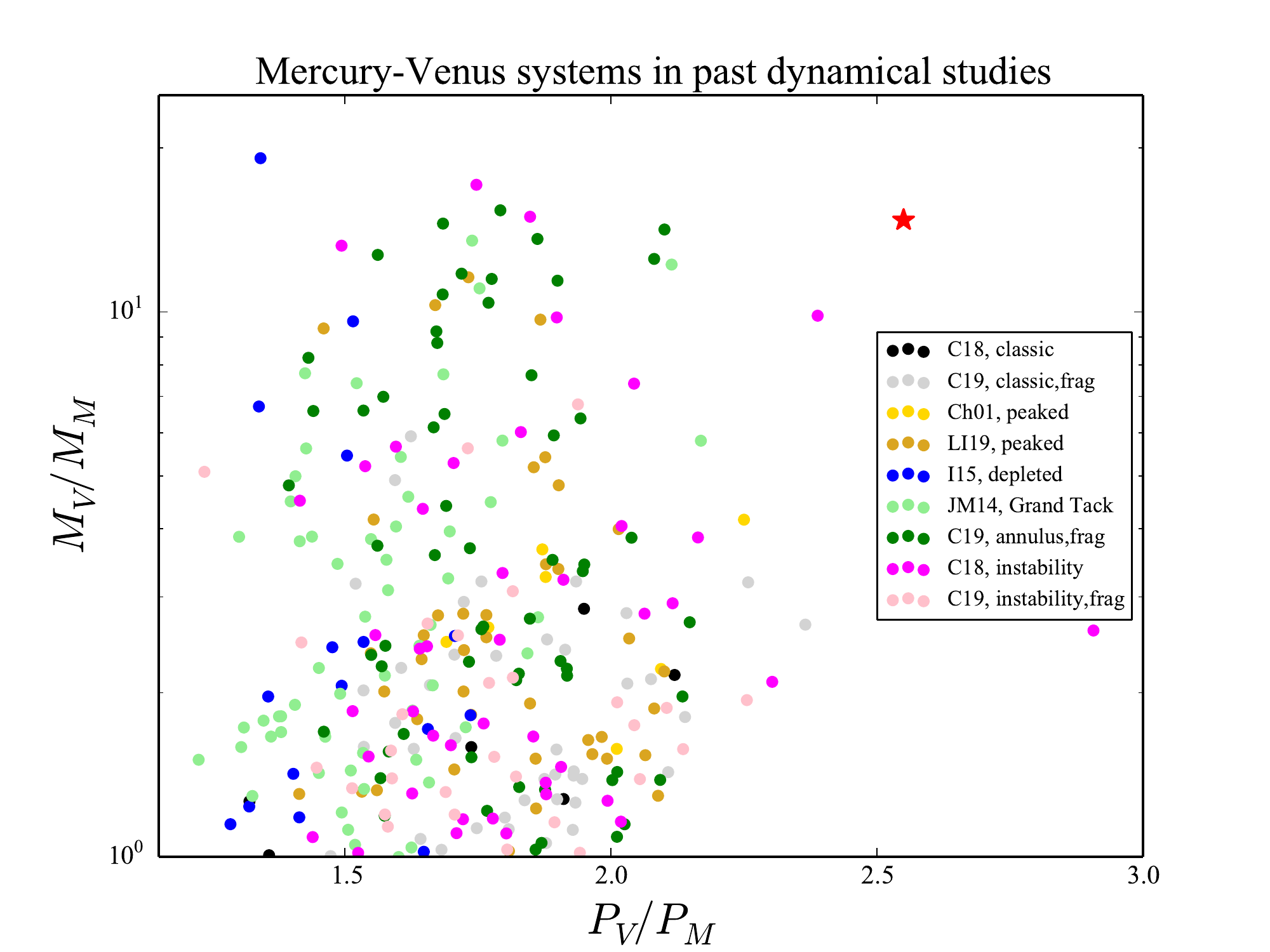}
	\caption{Summary of Mercury-Venus system mass and orbital period ratios in past dynamical studies.  For the purposes of this plot, we simply define a Mercury-Venus system as any simulation finishing with exactly two planets inside of 0.85 au.  The red star denotes the solar system value, and the different colored points correspond to simulation sets from the literature assuming different initial conditions.  The acronyms are as follows: C18: \citet{clement18}, C19: \citet{clement18_frag}, Ch01: \citet{chambers01}, LI19: \citet{lykawka19}, I15: \citet{izidoro15}, JM14: \citet{jacobson14}.  We direct the reader to the text in section \ref{sect:motivation} for a more detailed explanation.}
	\label{fig:new_fig}
\end{figure*}

Mercury's high mean density \citep{hauck13,nittler17} and depleted volatile inventory \citep{nittler11} have long been interpreted to imply that much of the mantle of a primordially massive proto-Mercury of roughly chondritic composition was eroded through a series of one or more violent impacts \citep{benz88}.  Such a sequence of events is particularly compelling as modern terrestrial planet formation models systematically struggle to produce planets with masses and orbits analogous to those of Mercury \citep[for recent reviews see:][]{ebel17,ray18_rev}.  Indeed, the Mercury analogs that do form in simulations assuming that the terrestrial-forming disk extended inward to $\sim$0.2-0.3 au are systematically too massive by around an order of magnitude \citep{chambers01,obrien06,lykawka17,lykawka19}.  Conversely, studies that assume the inner solar system grew from a narrow annulus of material \citep{wetherill78,agnor99,morishima08,hansen09,ray17sci,clement18_frag} occasionally generate diminutive planets interior to Venus when an embryo is scattered out of the disk.  In this manner, multiple authors have investigated how chemical \citep[e.g.: a fossilized silicate snowline as proposed in ][]{morby16_ice} or dynamical processes \citep{batygin15,ray16} might have played a role in truncating the inner terrestrial disk around $\sim$0.7 au.  However, it is still difficult to self-consistently reconcile Mercury's precise dynamics in these models as the resulting analog planets inhabit orbits that interact strongly with Venus \citep{lykawka19}, and the modern Mercury-Venus period ratio ($P_{V}/P_{M,ss}=$ 2.6) is almost never reproduced \citep{clement19_merc}.  We summarize these shortcomings within the contemporary terrestrial planet formation literature in figure \ref{fig:new_fig} and the accompanying caption.

While a collisional origin for Mercury is a compelling explanation for its unique composition, the impact geometry required in a scenario where proto-Mercury strikes a more massive body \citep[presumably proto-Venus, requiring collisional velocities as high as $\sim$3 times the mutual escape velocity:][]{asphaug14,jackson18,chau18} are fairly uncommon in terrestrial planet formation simulations \citep{clement19_merc}.  Moreover, Venus' lack of an internally generated dynamo and natural satellite has been proposed to suggest that its accretion was never interrupted by such an event \citep{jacobson17b}.  Additionally, it is unclear how Mercury's aphelion might be driven away from Venus' orbit after this violent collisional encounter.  Indeed, eventual re-accretion by Venus or collision with the Sun are far more likely outcomes, even for a range of possible fragment ejection velocities \citep{clement19_merc}.

It is also possible that proto-Mercury was repeatedly struck by a smaller object \citep{benz07}.  Given the appropriate selection of a target-projectile mass ratio, such a series of erosive hit-and-run impacts \citep[e.g.:][]{asphaug06} can involve less-extreme mutual velocities while still falling in to the catastrophically destructive collisional regime \citep[e.g.:][]{leinhardt12,stewart12,movshovitz16,gabriel20}.  Through this process, Mercury re-accretes from the disrupted material in a manner such that it attains a high mean density.  The remainder of the disrupted primordial mantle in this scenario would be removed by other processes \citep[e.g.: PR drag causing cm-scale fragments to spiral in to the Sun, the removal of $\sim$km-scale debris via Yarkovsky drift, or interactions with the strong solar wind in the young solar system:][]{vokrouhlick00,melis12,spalding20} in order to prevent it from eventually falling back on to Mercury \citep{gladman09}.

This manuscript is the first in a sequence of papers dedicated to developing a functional dynamical model for Mercury's genesis. In the first few investigations of the series, we reexamine the Mercury problem from the ground up by presenting simple models capable of replicating the precise Mercury-Venus system.  In a forthcoming study, we plan to further develop and robustly interrogate these various scenarios within the larger context of dynamical evolution (i.e.: giant planet migration) and planet accretion in the solar system.  In this paper, we investigate the possibility that Mercury emerged as the lone survivor of a destabilized primordial generation of short-period proto-planets in the inner solar system.

The general motivation for our study is twofold.  First, our proposed scenario is similar to the evolutionary sequence envisioned by \citet{volk15}.  The authors speculated that the solar system might have once possessed a system of three or more $\gtrsim$Earth-mass planets in the vicinity of Mercury's modern orbit similar to that of known multi-planet systems such as Kepler-11 \citep{lissauer13}.  As \citet{volk15} found analogs of several known short-period Kepler systems with multiple planets to be quasi-stable, the authors proposed that the solar system's hypothetical primordial population of short-period planets was lost in an instability that left behind Mercury as a relic, while the other terrestrial planets' orbits remained largely undisturbed.  However, there are several major shortcomings of the scenario originally envisioned in \citet{volk15}.  We expound on these important caveats, and discuss the connection between their work and ours in section \ref{sect:motivation}.. The secondary motivation for our current work originates from recent high-resolution studies of planetesimal accretion and runaway growth within the gas disk phase \citep{carter15,walsh19,clement20_psj,woo21}.  These studies find runaway growth to be highly efficient at generating quasi-stable chains of $\sim$0.1-0.4$M_{\oplus}$ embryos in the innermost sections of the terrestrial disk.  Thus, our work envisions a less-extreme version of the \citet{volk15} scenario; wherein a system of 3-6 $\sim$Mars-mass embryos extends from $\simeq$0.2 au to just inside of Venus' orbit.  We test our hypothesis with a suite of numerical simulations utilizing a calculation method that accounts for the fragmentation of colliding objects \citep{chambers99,chambers13}, and consider a range of possible mass and orbital distributions for the primordial inner solar system.  The masses of our hypothetical proto-planets \citep[similar to the modern masses of Mercury and Mars, as opposed to the $\sim$1 $M_{\oplus}$ objects proposed by][]{volk15} are loosely consistent with isotopic evidence \citep{budde16} indicating that the inner solar system was relatively starved of material from beyond Jupiter's orbit around 2 Myr after the formation of CAIs (Calcium Aluminum-rich Inclusions).  While the reason for this isolation of reservoirs is still debated (either the result of Jupiter's growth: \citealt{kruijer17}, or the consequence an intrinsic disk feature: \citealt{brasser20_nat}), it is likely that the inner solar system's evolution bifurcated from that of super-Earth hosting stars rather early \citep[e.g.:][]{izidoro17,lambrechts19}.

\section{Methods}
\label{sect:meth}

\subsection{Motivation}
\label{sect:motivation}

Both the isotopic record derived from meteorites \citep{budde16,kruijer17,brasser20_nat} and the predictions of modern disk models \citep[e.g.:][]{morby12_peb,levison15,johansen15,draz16,lambrechts19,bitsch19} rather strongly disfavor super-Earth-formation having been achieved in the inner parts of the solar nebula.  This staunchly contradicts the original model of \citet{volk15}.  Moreover, it is difficult to reconcile the high inferred occurrence frequencies of super-Earths \citep[$\sim$50$\%$ or higher around Sun-like stars:][]{chiang13,zhu18} with a scenario where such a chain of planets in the solar system is cataclysmically ground to dust.  Indeed, the orbital period ratios of such multi-planet systems seem to imply that their architectures were reshaped by delayed epochs of giant impacts and instability \citep{pu_wu15,izidoro17,izidoro19}. Thus, it is unclear why, unlike the vast majority of know systems, a hypothetical population of super-Earths in the solar system would be catastrophically demolished.  Indeed, recent work in \citet{poon20} using a collisional fragmentation algorithm concluded that the formative episodes of giant impacts and instabilities within chains of hot super-Earths typically do not yield collisional velocities in the cataclysmically destructive regime \citep[e.g.:][]{kokubo10,genda12,leinhardt12,stewart12,gabriel20}.  For these reasons, we intentionally deviate from the original proposition \citet{volk15}, and remind the reader that our work does not invoke super-Earth formation in the solar system, or exotic dynamical processes (e.g.: planetesimal shepherding) to generate our proposed initial conditions \citep[see additional discussion in:][]{ray16,deienno20,lenz20,bean20}.

Figure \ref{fig:new_fig} summarizes the systematic inability of dynamical models to generate Mercury-Venus systems.  For simplicity, we define a ``Mercury-Venus'' system as any numerical simulation finishing with exactly two planets inside of 0.85 au with $M_{M}<M_{V}$.  The black points illustrate a set of 100 simulations from \citet{clement18} assuming the so-called classic initial conditions \citep[$\sim$5 $M_{\oplus}$ of terrestrial planet forming material distributed uniformly between 0.5-4.0 au, e.g:][]{chambers98,ray09a}.  Here, Mercury forms directly within the massive disk in the same manner as Earth and Venus.   The grey points depict a batch of 100 simulations from \citet{clement18_frag} assuming the same classic initial conditions and also including a prescription \citep{chambers13} accounting for the effects of collisional fragmentation.  In successful realizations, Mercury forms small as the result of a series of imperfect accretion events.  The gold points represent a suite of 16 simulations reported in \citet{chambers01} where the Mercury forming region is modeled with a linearly increasing surface density of embryos and planetesimals between 0.3-0.7 au (hence the disk is ``peaked'').  Successful outcomes occur when Mercury forms directly from within this mass-depleted inner disk component.  The tan points plot 40 Mercury-Venus analog systems formed in 89 simulations in \citet{lykawka19} considering similar inner disk components.  The work of \citet{izidoro15}, comprising 60 numerical simulations where the slope of the disk surface density profile is varied, is denoted with blue points.  Successful Mercury analogs are produced when an embryo is scattered out of the disk.  The light green points illustrate a sample of 71 Mercury-Venus systems formed in 118 total simulations of terrestrial planet formation in the Grand Tack scenario \citep{walsh11} from \citet{jacobson14}.  As in the depleted asteroid belt framework of \citet{izidoro15}, successful Mercury analogs are typically embryos that scatter and dynamically decouple from the disk.  The green points plot a set of 125 simulations in \citet{clement18_frag} including a fragmentation prescription where the terrestrial planets form out of a narrow annulus of material ($\sim$2 $M_{\oplus}$) between 0.7-1.0 au as envisioned in \citet{hansen09}.  Successful Mercury analogs are occasionally generated when an embryo is scattered out of the annulus.  Conversely, the magenta points depict a subset of Mercury-Venus systems generated in 800 simulations from \citet{clement18} where the giant planet instability \citep{Tsi05,nesvorny12} occurs within the first 10 Myr of the process terrestrial planet formation.  Appropriate Mercury analogs are produced when the giant planets' resonant perturbations liberate an embryo from the planet-forming disk and strand it on a Mercury-like orbits.  Finally, the pink points represent set of 600 simulations from \citet{clement18_frag} modeling the same early instability scenario and also incorporating the effects of collisional fragmentation.

Given the trends depicted in figure \ref{fig:new_fig}, we begin our study with a simple question that will, unfortunately, remain unanswered in our manuscript: \textit{What was the structure of the terrestrial disk in the Mercury-forming region?}  Perhaps the disk conditions that gave rise to Mercury are still completely unknown \citep[e.g.:][]{bean20}, and the planets' origin will remain a mystery until future exploration of the surface shines light on the nature of the Mercury-forming planetesimals, and detailed disk observations and models conclusively reveal the elusive initial conditions that generate Mercury-like planets.  It is also possible that the standard setups employed by contemporary formation studies are qualitatively correct \citep[for example:][]{walsh19,lykawka19,clement18_frag,deienno19}, and the apparent inability of such models to reproduce Mercury (figure \ref{fig:new_fig}) is the result of a flaw in methodology or approach (i.e.: insufficient resolution, inappropriate collisional treatments, etc.).  Regardless of which proposition (or both) is correct, we argue that the scope of the Mercury problem permits some margin for experimentation and exploration outside of the conventional bulwark of solar system evolutionary narratives.

In this paper we investigate a scenario where a quasi-stable chain of 4-6 large, well-spaced embryos emerge in the region interior to the terrestrial forming disk.  While these initial conditions might seem largely contrived, we believe that they are not particularly fictitious given the diversity of disk structures generated in hydrodynamical and N-body models.  For instance, \citet{draz16} developed a semi-analytical disk model to follow the evolution of dust and pebbles in the Sun's gaseous nebula and found that planetesimals form and pile-up in a narrow annulus of width $\sim$ 0.3-3.0 au beginning around $a \sim$0.3-1.0 au.  Thus, it is important to refrain from grounding any investigation in a narrow assumption that planetesimal growth and, by extension, planet formation transpired identically throughout the terrestrial-forming disk.  This is particularly relevant given the tendency of disk models to find planetesimal formation to be highly localized and sensitive to variations in initial conditions \citep[e.g.:][]{levison15,draz16,yang17,draz18,lenz19,lenz20}.  Additionally, the conversion of the planetesimals that do form into large embryos \citep[similar to those studied in this paper:][]{koko_ida_96,koko_ida00,chambers06} occurs with similarly variable efficiencies at different locations in the disk \citep{morishima08,walsh19,wallace19}.  In particular, high-resolution N-body simulations of this epoch in \citet{clement20_psj} found the innermost regions of the terrestrial-forming disk to be highly processed after the runaway growth phase; and virtually devoid of planetesimals near the Mercury-forming region (bins of $\simeq$0.45-0.6 au in that work).  Moreover, it seems plausible to envision a population of embryos traversing the disk under the influence of Type I migration becoming stranded in the Mercury-forming region during the disk's photo-evaporation phase.  Indeed, new work in \citet{woo21} considering a variety of conceivable disk parameters and giant planet configurations reveal similar chains of dynamically isolated embryos (although their dynamical separations are not as broad in terms of mutual Hill radii as presumed in our simulations; see figures 2 and 12 in that work).

\subsection{Numerical Simulations}

We select a modified version of the $Mercury6$ Hybrid integrator \citep{chambers99} described in detail in \citet{chambers13} for our numerical simulations.  This augmented package includes algorithms designed to incorporate the various fragmentation regimes mapped in \citet{leinhardt12} and \citet{stewart12}.  In short, when an erosive collision is detected, the code calculates the mass of the largest remaining remnant, and divides the remaining material into a number of equal-mass fragments that are ejected at $v \simeq 1.05 v_{esc}$ in uniformly spaced directions in the collisional plane.  While this selection of ejection velocity is not precisely physically motivated, it represents a reasonable compromise between the premature re-accretion of displaced material and the creation of excessively dynamically excited fragments \citep[see][for a recent study that tested different ejection velocities]{clement19_merc}.  The total number of fragments is determined by the user-established minimum fragment mass (MFM); which we set to 10$\%$ that of Mercury's modern mass (0.0055 $M_{\oplus}$) in our simulations.  Establishing a MFM mass is necessary to avoid injecting an unreasonable number of particles in a simulation such that the calculation is no longer computationally feasible.

As the parameter space of possible primordial configurations of short-period proto-planets is extensive, we limit our current study to systems of equal-mass objects in the vicinity of Mercury's modern orbit.  In four independent batches of integrations (table \ref{table:ics}) we investigate systems of $M_{emb}=$ 0.3, 0.1, 0.05 and 0.025 $M_{\oplus}$ embryos, distributed approximately uniformly between 0.25-0.6 au (we impose a random variation of $\delta_{a} \lesssim$ 0.02 au on each semi-major axis to generate different initial conditions).  Eccentricities and inclinations for the embryos are selected at random from nearly circular, co-planar Rayleigh distributions ($\sigma_{e}=$ 0.01, $\sigma_{i}=$ 0.1$^{\circ}$), and the remaining angular orbital elements are assigned randomly by sampling from uniform sets of angles.  While our initial conditions are clearly somewhat contrived in the absence of high resolution dynamical investigations of runaway growth at small radial distances (e.g.: section \ref{sect:motivation}), our higher values of $M_{emb}$ (0.1 and 0.3 $M_{\oplus}$) are loosely based off recent high-$N$ studies of planetesimal accretion in the gas disk phase \citep{walsh19,clement20_psj,woo21}.  Contrarily, our lower presumed values of $M_{emb}$ (0.05 and 0.025 $M_{\oplus}$) are selected to boost the probability of forming Mercury analogs with the correct mass (i.e.: they are not particularly physically motivated beyond the fact that they are similar to the actual mass of Mercury).  It is also important to note that, in all cases, our embryos begin the simulation separated from one another by $\sim$30-50 mutual Hill radii ($R_{h}$).  Thus, our initial conditions are selected such that the system of short-period planets should possess a relatively high degree of Hill stability \citep[e.g.:][]{chambers96,koko_ida_98,petit18} in the absence of perturbations from the other planets.

In all of our simulations, we include Venus, Earth, Mars, Jupiter and Saturn on their modern orbits.  As Venus typically accretes at least one of the embryos in our simulations (section \ref{sect:venus}), we reduce its initial mass to $M_{Venus}=$ 0.6 $M_{\oplus}$ in our two sets testing the highest values of $M_{emb}$ in order to boost the probability that Venus finishes with the correct mass.  Each of our systems is integrated for 100 Myr utilizing a 1.5 day time-step and algorithms designed to account for the effects of relativity.  

\begin{figure}
	\centering
	\includegraphics[width=.5\textwidth]{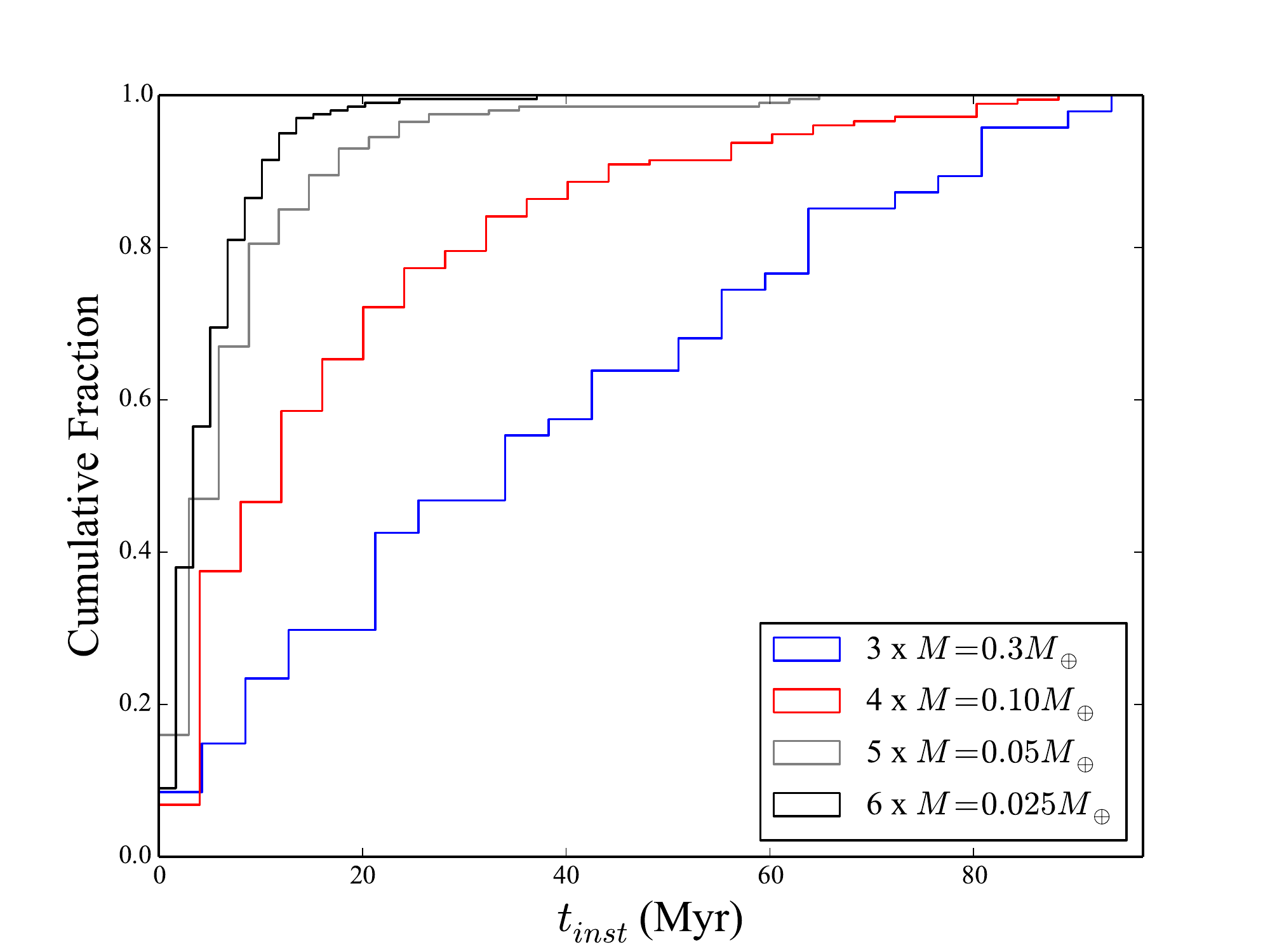}
	\caption{Cumulative fraction of instability times for our various simulations batches (only systems that experience instabilities are included in this plot).}
	\label{fig:tinst}
\end{figure}

The fraction of simulations that experience an instability resulting in the loss of one or more proto-planets is given in table \ref{table:ics}.  For simplicity, we define the ``instability time'' as the first collisional encounter involving an embryo.  Instabilities typically transpire within a few tens of Myr in simulations evaluating lower values of $M_{emb}$ (figure \ref{fig:tinst}), and occur with at lower frequencies and after longer delays in our integrations investigating $M_{emb}=$ 0.3 and 0.1 $M_{\oplus}$.  As our various configurations of proto-planet masses begin our integrations with roughly equivalent mutual Hill spacings ($\sim$30-50 $R_{h}$), the lengthy instability delays in our $M_{emb}=$ 0.3 $M_{\oplus}$ batch are somewhat conspicuous \citep[e.g.:][]{chambers96}.  We find that this higher degree of stability in our more massive embryo configurations is a result of the larger proto-planets being initialized on more widely separated orbits.  We expound further on this discrepancy in section \ref{sect:destab}.

\begin{table*}
	\centering
\begin{tabular}{c c c c c c}
\hline
Set  & $N_{emb}$ &  $M_{emb}$ ($M_{\oplus}$) & $M_{Venus}$ ($M_{\oplus}$) & $N_{sim}$ & $N_{instb}/N_{sim}$ ($\%$) \\
\hline
1 & 3 & 0.3 & 0.6 & 200 & 31\\
2 & 4 & 0.1 & 0.6 & 200 & 88\\
3 & 5 & 0.05 & 0.815 & 200 & 100\\
4 & 6 & 0.025 & 0.815 & 200 & 100\\
\hline
\end{tabular}
\caption{Summary of initial conditions for our various simulations.  The columns are as follows: (1) The simulation set, (2) the total number of embryos placed interior to Venus' orbit, (3) the masses of the individual embryos, (4) the initial mass of Venus, (5) the total number of simulations, and (6) the percentage of all simulations that experience an instability.}
\label{table:ics}
\end{table*}

\subsection{Core Mass Fraction calculation}

We evaluate the compositional evolution of the objects in each simulation using the methodology of \citet{chambers13} for a first order approximation of the final Mercury-analogs' Fe/Si ratios.  We begin by assuming that each inner planet's initial composition is approximately chondritic, and fully differentiated such that 30$\%$ of the total mass is in an iron-rich core and the silicate mantle makes up the remaining 70$\%$.  When an erosive collision occurs between two bodies in the simulation, fragments are first generated from the mantle of the projectile, then its core, then the mantle of the target body, and finally from the target's (largest remnant) core material.  

An alternative methodology for assembling the final planets' bulk compositions would be to assign each fragment with an equal mixture of the cumulative ejected material.  One could imagine this as a crude approximation of the fragments themselves having re-accreted from a sea of vaporized material.  However, we experimented with such an implementation and determined that it did not significantly alter the final CMF distributions in the majority of our systems.  This is because collisional fragments are rarely involved in prolonged, subsequent episodes of CMF-altering collisions.  While such fragments do occasionally experience repeated follow-on hit-and-run events with the larger remnant particle, such interactions almost exclusively result in the eventual accretion of the fragment with no net transfer of mantle or core material outside of the interacting system.  We elaborate further on whether or not our fragmentation algorithm is an accurate physical representation of CMF-alteration during erosive collisions in section \ref{sect:merc}.

The pitfall of our chosen CMF-tracking implementation is the obvious degeneracy that arises when both the mantle and core of an object is eroded.  In this case, the various fragments are essentially randomly assigned CMFs of either 0.0 or 1.0, with one fragment typically possessing a mixture of the two to ensure mass convergence.  If one of these fragments survives the simulation and becomes a Mercury-analog, the simulation's success or failure in terms of Mercury's mean density would be determined by what material was randomly assigned to the individual fragment that went on to become Mercury.  While we find that this situation rarely arises as the majority of fragments are re-accreted by either the remnant particle or another embryo in the system (and seldom become Mercury analogs themselves), we still focus the majority of our analysis in the subsequent sections on our various simulation sets' collective statistical properties.  However, we also discuss individual systems that are excellent solar system analogs as examples of how our proposed scenario might have transpired.

\section{Results}

\subsection{Destabilization mechanism}
\label{sect:destab}

\begin{figure*}
	\centering
	\includegraphics[width=.95\textwidth]{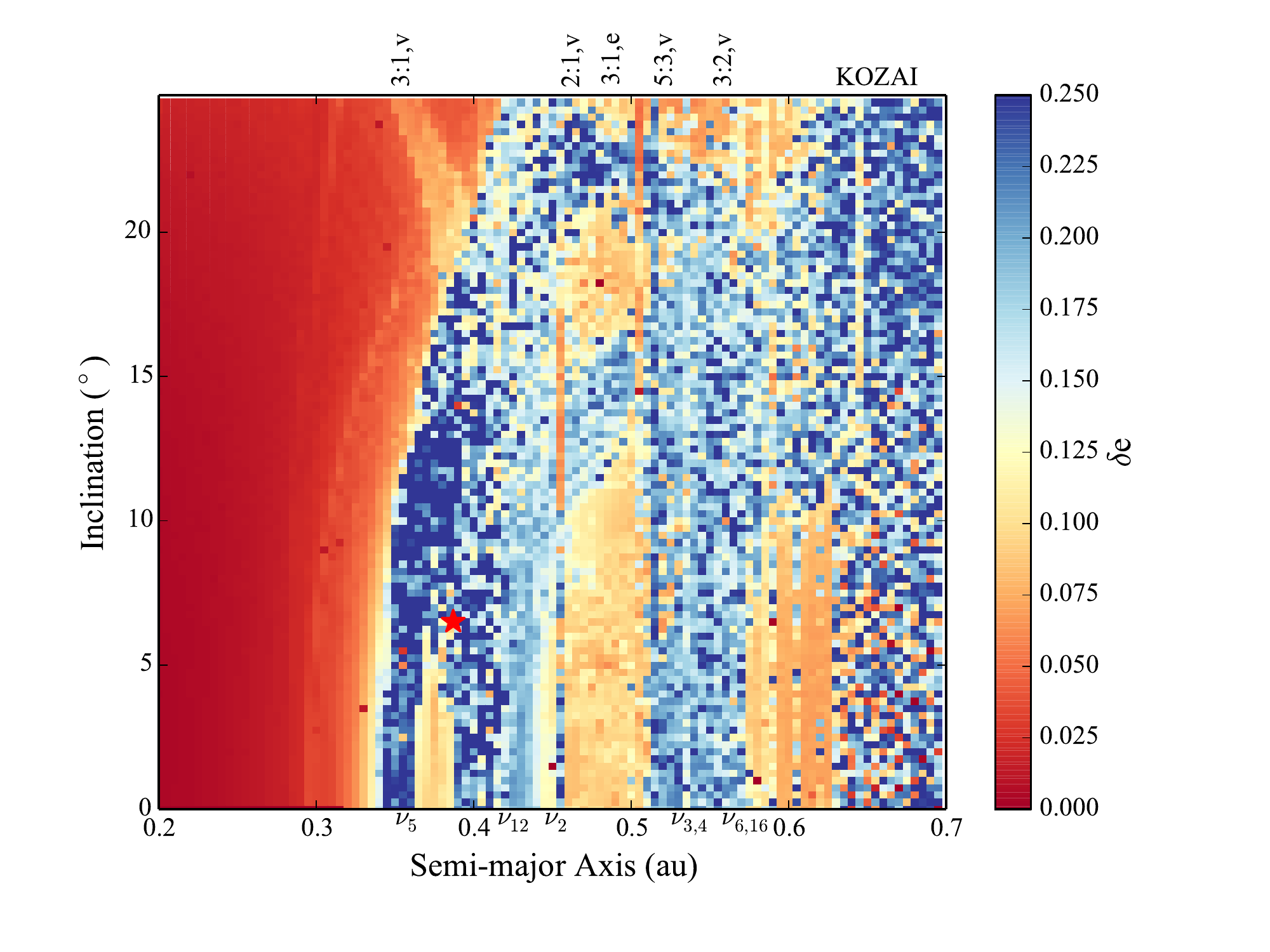}
	\caption{Maximum change in eccentricity of Mercury on various orbits in the inner solar system.  The map is generated by performing 10,000 simulations of the modern solar system with the outer seven planets at their current orbital locations. The position of Mercury is varied in $a/i$ space in each simulation. The color of each point corresponds to half the maximum change in the Mercury's eccentricity over a 10 Myr integration.  The locations of the dominant first and second order MMRs with the terrestrial planets are denoted on the top of the plot.  The most significant secular resonances are identified by their respective eccentricity-averaged, zero-inclination locations \citep[see][]{michel97} on the bottom of the plot.  As the locations of $\nu_{3}$ and $\nu_{4}$, as well as $\nu_{6}$ and $\nu_{16}$, are so close together we identify each pair of resonances as $\nu_{3,4}$ and $\nu_{6,16}$, respectively.  The large red star denotes Mercury's modern orbit.}
	\label{fig:map}
\end{figure*}

The region of the modern solar system interior to $\simeq$0.65 au is quite complex dynamically \citep[outside of 0.65 au, in the vicinity of Earth and Venus, NEA (Near Earth Asteroids) dynamics are primarily governed by Kozai cycles:][]{michel96}.  Overlapping mean motion (MMR) and secular resonances in the region result in large swaths of parameter space where our hypothetical proto-planets' orbits evolve chaotically \citep[e.g.: figure \ref{fig:map}, see also:][]{michel97,laskar09,batygin15b}.  Most significant among these various unstable zones, objects with semi-major axes near the intersection of the 3:1 MMR with Venus and the powerful $\nu_{5}$ secular resonance at $\sim$0.35 au (leftmost blue regime in figure \ref{fig:map}) are highly unstable, and can easily be excited in eccentricity to the point of loss or collision with another proto-planet in relatively short timescales.  Orbits around $a=$ 0.46 and $a=$ 0.55 au are also unstable in our simulations (in that objects attain Venus-crossing orbits within 10 Myr); a consequence of overlaps between $\nu_{2}$, $\nu_{12}$, and the 2:1 with Venus, and the $\nu_{16}$ $\nu_{6}$, $\nu_{3}$, $\nu_{4}$, and 3:2 MMR, respectively.

\begin{figure}
	\centering
	\includegraphics[width=.5\textwidth]{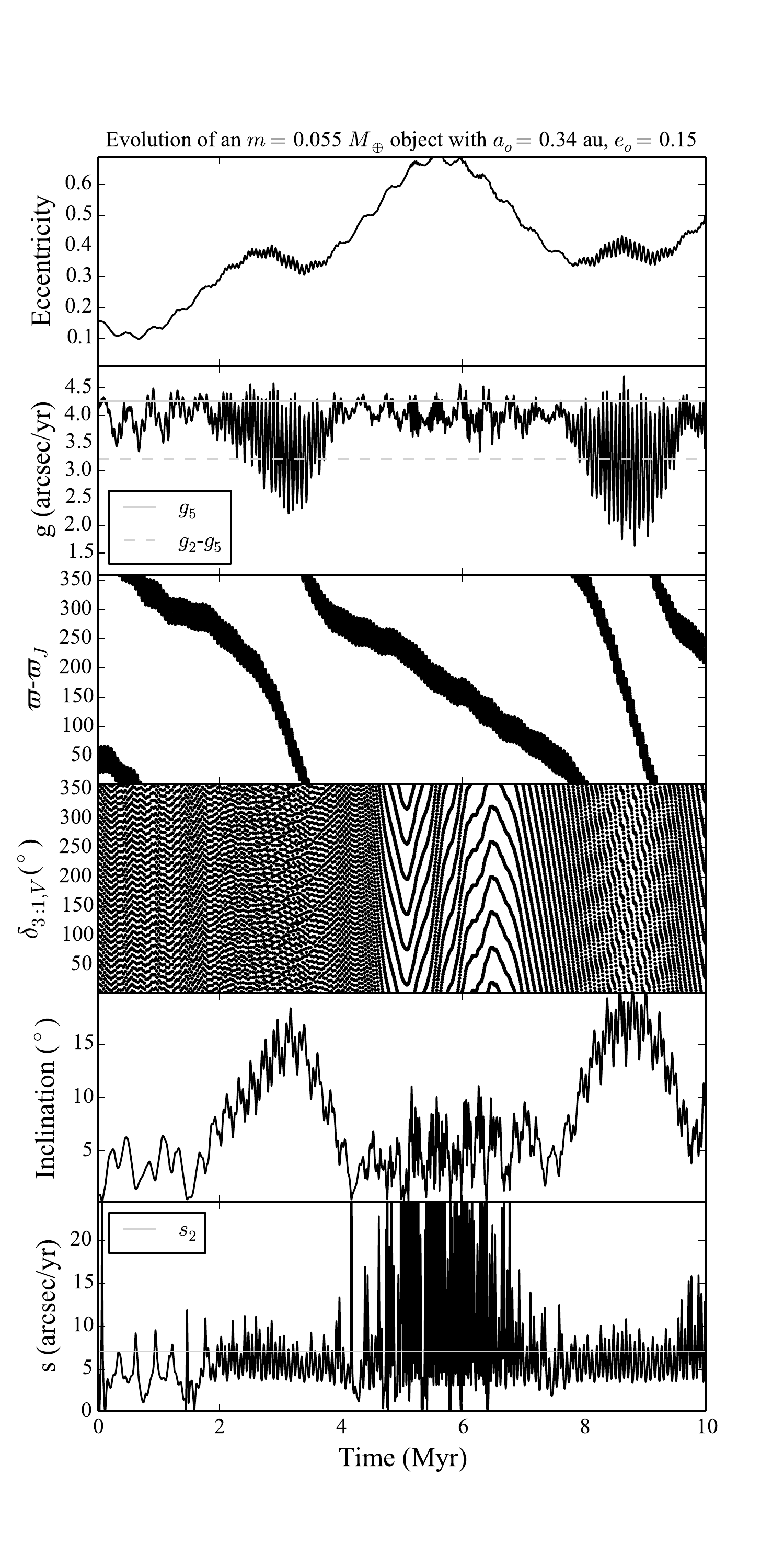}
	\caption{10 Myr integration of the solar system where Mercury is placed at the intersection of the $\nu_{5}$ secular resonance and Venus' interior 3:1 MMR.  The top panel depicts Mercury's eccentricity evolution, while the second and third panels (Mercury's rate of perihelion precession: $g$; and a critical angle of the $\nu_{5}$ resonance: $\varpi - \varpi_{J}$)  demonstrate how the solar system's fifth eccentric eigenfrequency, $g_{5}$, drives episodes of large oscillations in the planets eccentricity.  The high eccentricity epochs between $\sim$4-8 Myr are also affected by the proximity to the 3:1 MMR with Venus, which is depicted by the behavior of the relevant resonant angle in panel four.  The fifth panel shows Mercury's inclination evolution, which is predominantly driven by the $\nu_{12}$ resonance as demonstrated by the evolution of the planet's nodal precession rate, $s$, in panel six.}
	\label{fig:test}
\end{figure}

Figure \ref{fig:test} demonstrates how perturbations from Jupiter's $g_{5}$ mode can easily excite objects with masses akin to those of our embryos on to crossing orbits \citep[see also:][]{batygin15b}.  The plot depicts a 10 Myr integration of the solar system \citep[utilizing the $Mercury6$ Hybrid integrator and a 1.5 day time step:][]{chambers99} where Mercury's semi-major axis is shifted to the unstable region in figure \ref{fig:map} at $a=$ 0.34 au.  It is important to recognize that the secular architecture of this system (or those of any of the other integrations presented in this manuscript) is not necessarily akin to that of the solar system.  For instance, Mercury's nodal precession rate induces a significant perturbation on Venus' inclination evolution in the modern solar system \citep[e.g.:][]{laskar90}, and thus the location and precise dynamics of the $\nu_{12}$ resonance is clearly different in our simulations than in the solar system by virtue of moving Mercury, or replacing it with a system of short-period proto-planets.  However, it is clear from figure's \ref{fig:map} and \ref{fig:test} that the influence of $g_{5}$ in the region interior to Mercury's modern orbit is quite signifiant in terms of its ability to excite the orbits of Moon-Mars-mass objects.  It is also important to note that this analysis only considers the giant planets on their modern orbits.  If our hypothetical epoch of short-period planet destabilization transpired in conjunction with the giant planet instability \citep[][discussed further in sections \ref{sect:analog} and \ref{sect:discuss}]{Tsi05,nesvorny12,deienno17}, it is plausible that the inward sweeping of $\nu_{5}$ towards its ultimate modern location might serve to destabilize our chains of embryos \citep[e.g.:][]{bras09,roig16}.

While it is beyond the scope of our present manuscript to fully scrutinize each individual instability and determine the precise destabilization mechanism, it is obvious that instabilities are fairly ubiquitous in our simulations (particularly for the lower-mass systems of embryos: table \ref{table:ics}).  Moreover, from a simple analysis of the encounter histories of our systems of proto-planets it is readily apparent that the highly perturbed region of orbital parameter space in the vicinity of the $\nu_{5}$ secular resonance plays a key role in driving our configurations towards an instability.  Indeed, $\gtrsim$45$\%$ of the first collisions in our simulations involve the embryo that originates closest to the 3:1 resonance with Venus.  Another $\sim$30$\%$ of our systems' instabilities unfold after the outermost embryo's eccentricity is gradually excited until the object inhabits a crossing orbit with Venus (at least partially as a consequence of being in close proximity to the chaotic region in figure \ref{fig:map} around the 3:2 MMR at 0.55 au).  Additionally, we used a set of preliminary runs to determine that placing the central embryo in our $M_{emb}=$ 0.3 $M_{\oplus}$ sets (the most resilient to destabilization; table \ref{table:ics}) at a semi-major axis roughly corresponding to the 3:1 MMR with Venus boosts the total fraction of systems experiencing an instability by $\sim$50$\%$.  

The highly chaotic orbital evolution of proto-planets in the radial regime interior to Mercury's modern semi-major axis (e.g.: figure \ref{fig:test}) is indeed somewhat conspicuous, and remarkably consistent with the hypothesis that the solar system once possessed additional planet-mass bodies interior to Venus \citep{volk15}.  The modern Mercury-Venus period ratio is peculiar in that it is quite larger than those of the other neighboring pairs of terrestrial planets, and is seldom matched in planet formation models \citep{clement19_merc}.  Furthermore, Mercury is positioned at the edge of one of the more regularly behaved islands in $a/i$ space (figure \ref{fig:map}) between Venus, and the $\nu_{5}$ resonance.  It may be impossible to ascertain whether or not additional planets formed inside of Venus' orbit.  However, it is easy to imagine a simplified version of our scenario, where an additional planet once existed between Mercury and Venus in the vicinity of the 3:2 MMR, $\nu_{16}$, $\nu_{3}$ overlap, and was destabilized and lost via a collision with Venus.  Further additional planets interior to Mercury might also have existed, and were eventually lost via dynamical interactions with the powerful $\nu_{5}$ resonance as discussed above.  Such hypothetical short-period planets might present an intriguing explanation for Mercury's peculiar composition if they underwent a series of erosive impacts with proto-Mercury en-route to being lost.  To ascertain the likelihood of this hypothetical scenario, we turn to our suites of full dynamical simulations (table \ref{table:ics}).

\subsection{Analog Systems}
\label{sect:analog}

\begin{figure*}
\centering
\includegraphics[width=.4\textwidth]{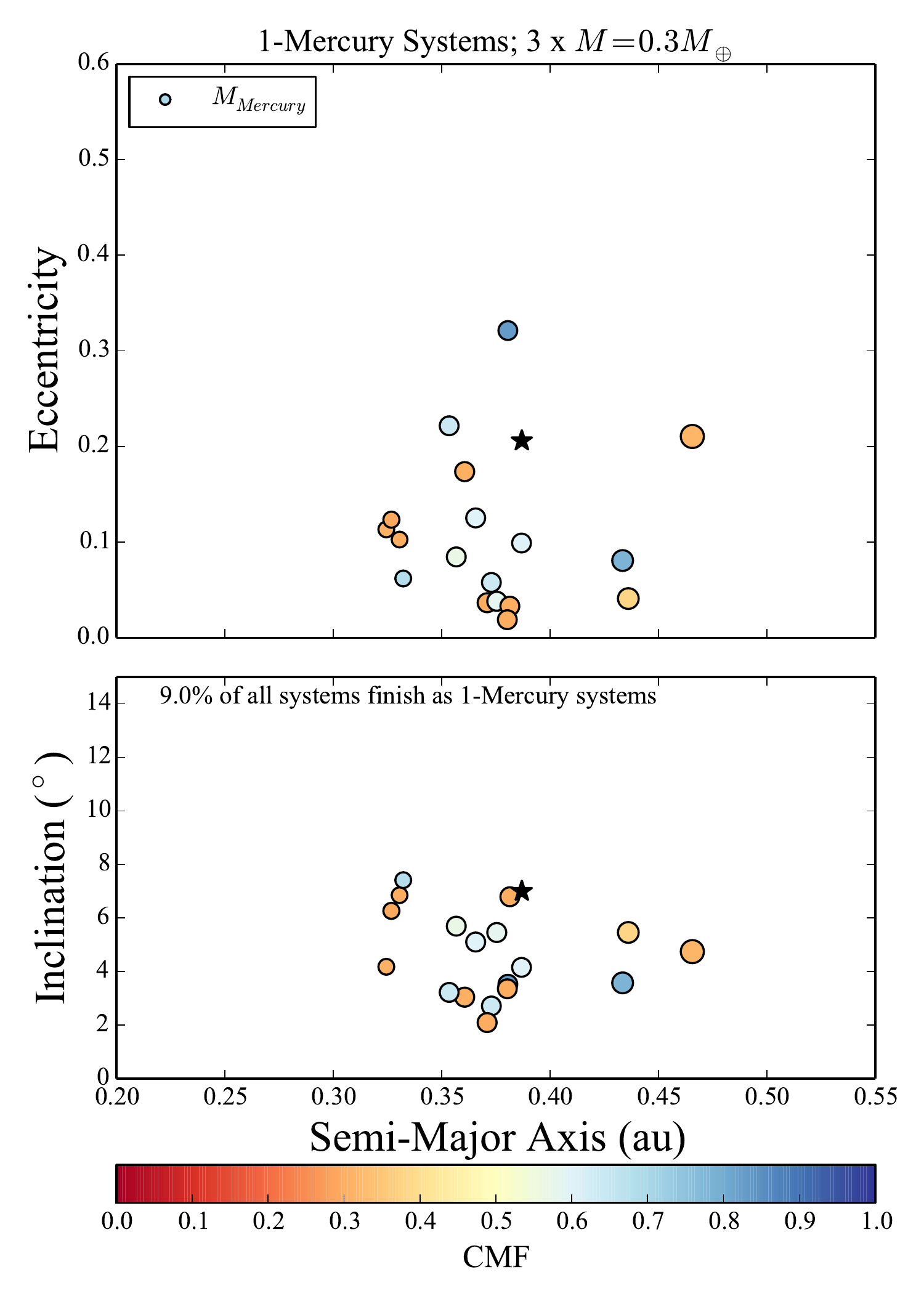}
\includegraphics[width=.4\textwidth]{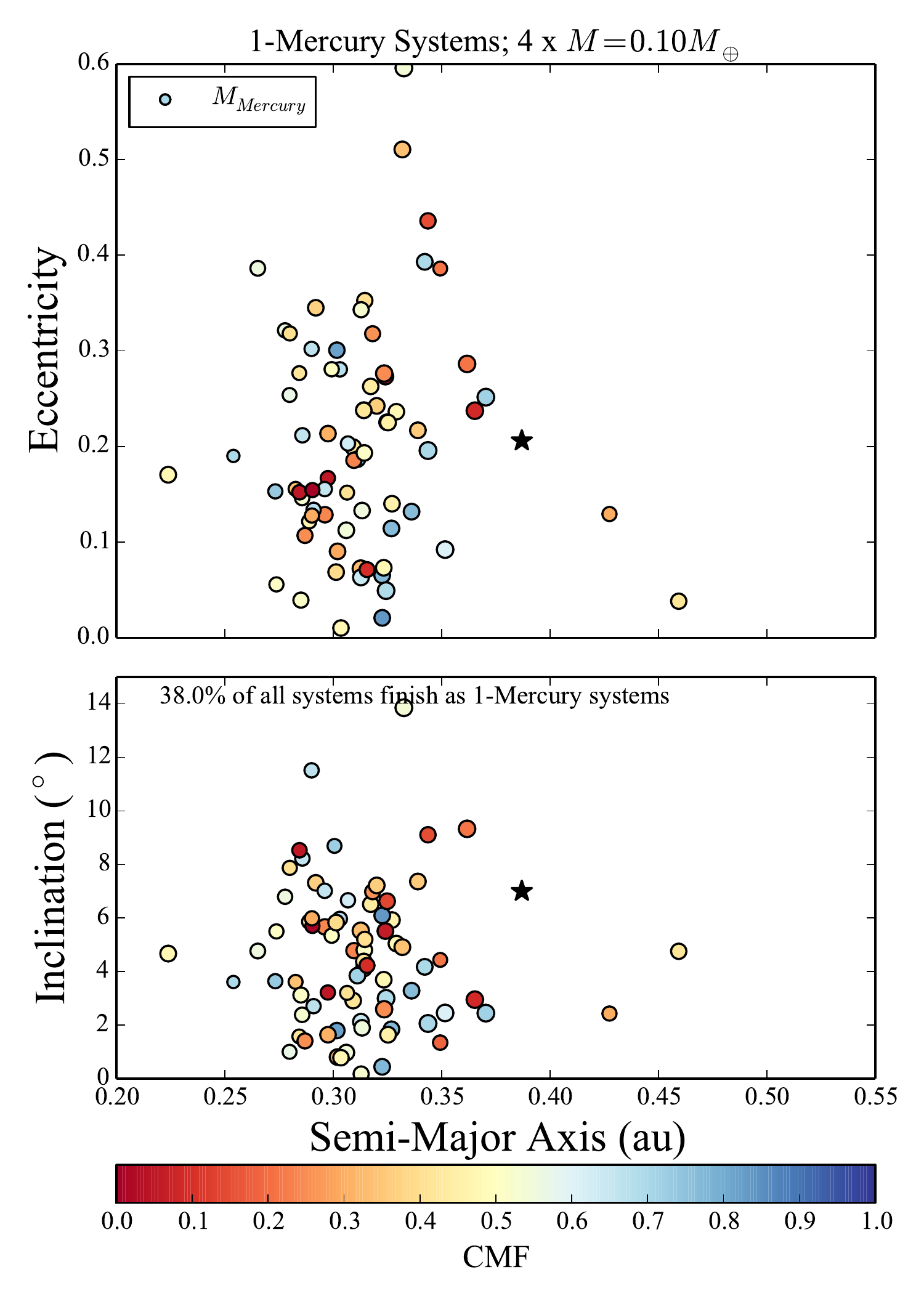}
\qquad
\includegraphics[width=.4\textwidth]{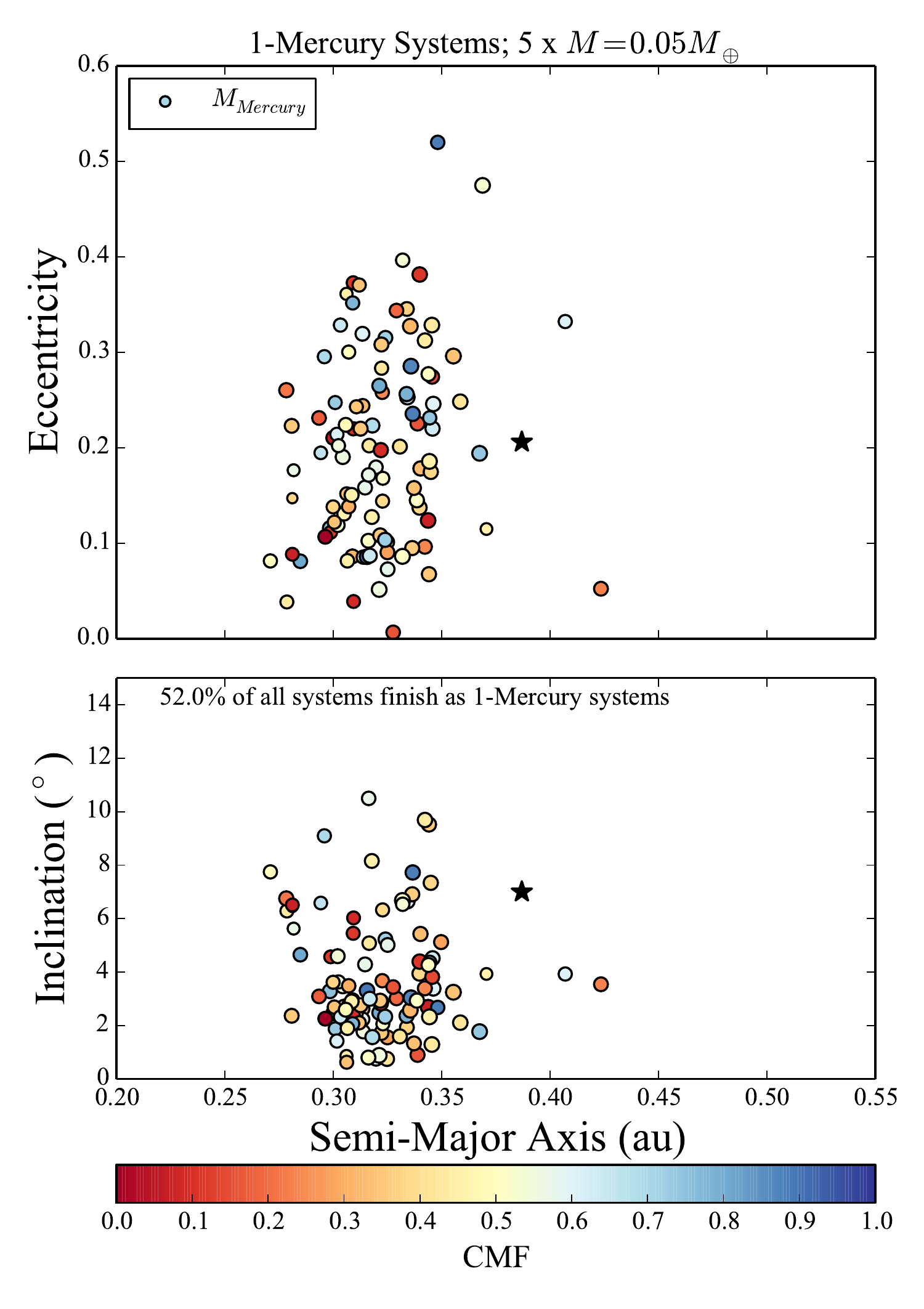}
\includegraphics[width=.4\textwidth]{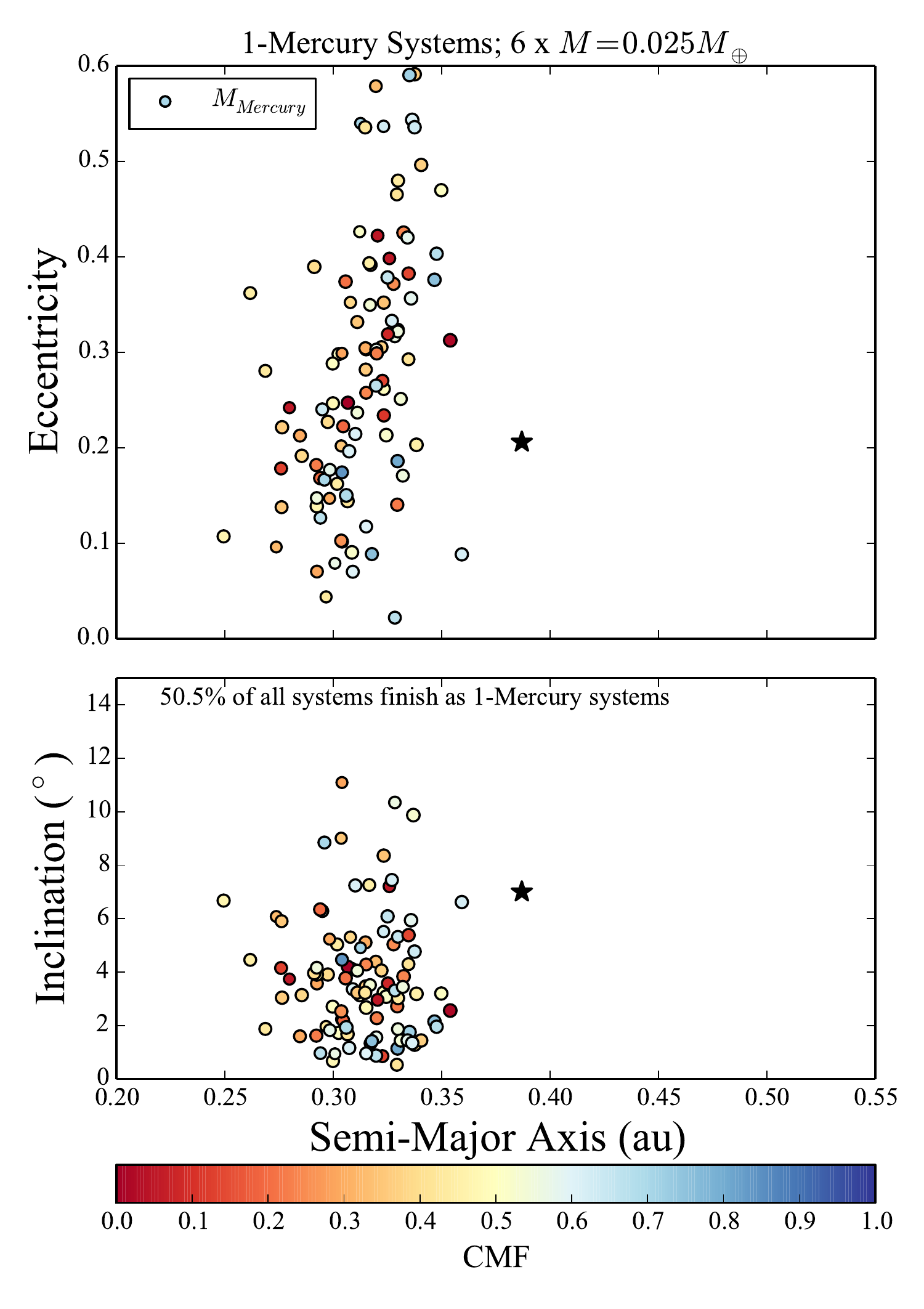}
\caption{Final orbits in our various batches of simulations (table \ref{table:ics}) for systems finishing with the correct number of planets (i.e.: one Mercury analog).  Each of the four panels depict a different simulation batch.  For each batch, the top sub-panel plots eccentricity vs. semi-major axis for each Mercury analog, and the bottom sub-panel displays the same systems in $a/i$ space.  The size of each point is proportional to the analog's mass (the mass of Mercury is plotted in grey in the upper left corner of each panel for reference), and the color of each point indicates the object's CMF.  Mercury's modern orbit is indicated in each sub-panel with a black star.}
\label{fig:mercs}
\end{figure*}

\begin{figure}
	\centering
	\includegraphics[width=.5\textwidth]{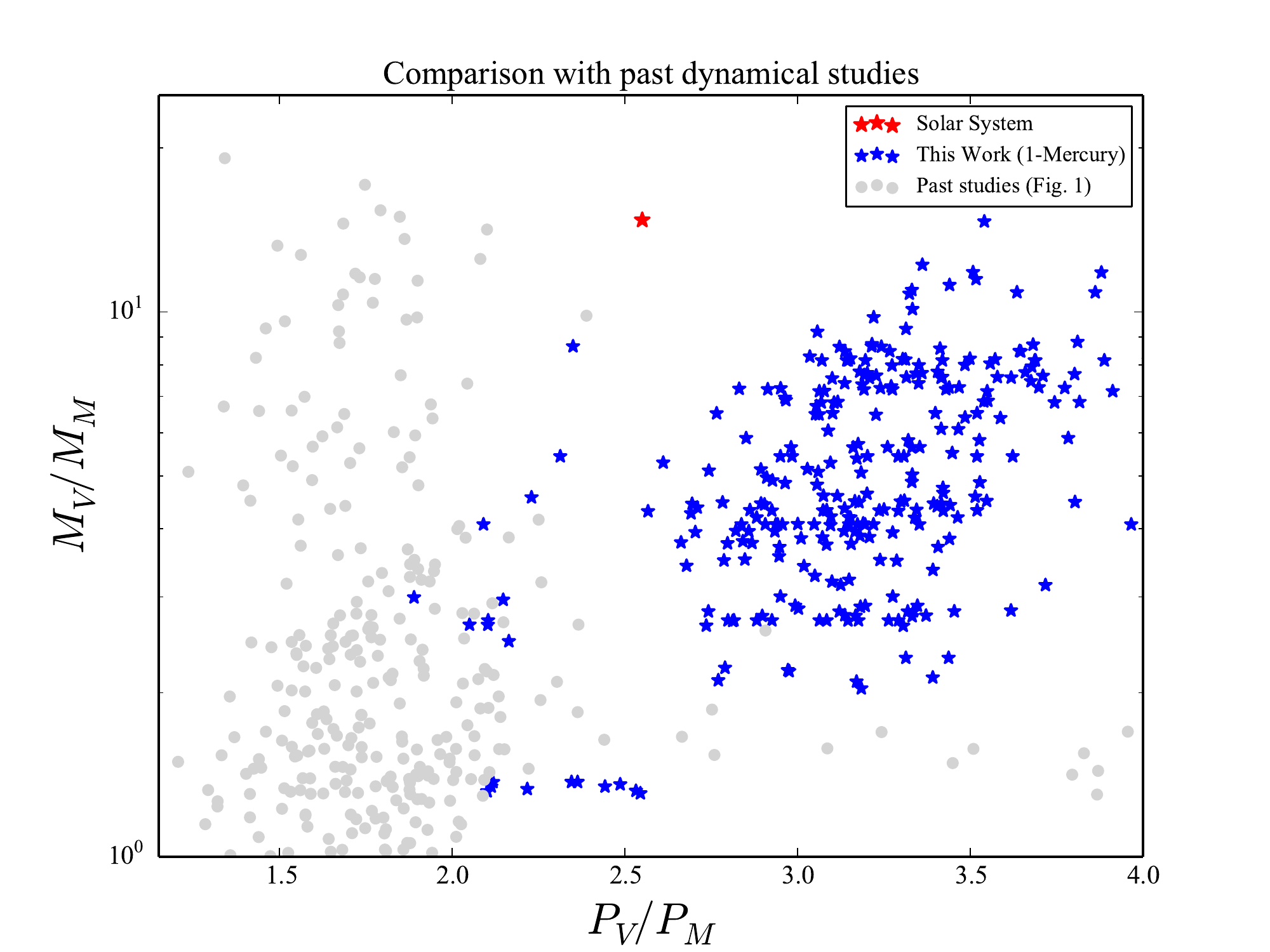}
	\caption{The same as figure \ref{fig:new_fig} except, here, the systems formed in past dynamical studies \citep{chambers01,izidoro15,clement18,clement18_frag} are plotted in grey and the 1-Mercury realizations from this work are depicted with blue stars.  The red star denotes the solar system values of $P_{V}/P_{M}$ and $M_{V}/M_{M}$.}
	\label{fig:new_fig_tw}	
\end{figure}

\begin{figure}
\centering
\includegraphics[width=.5\textwidth]{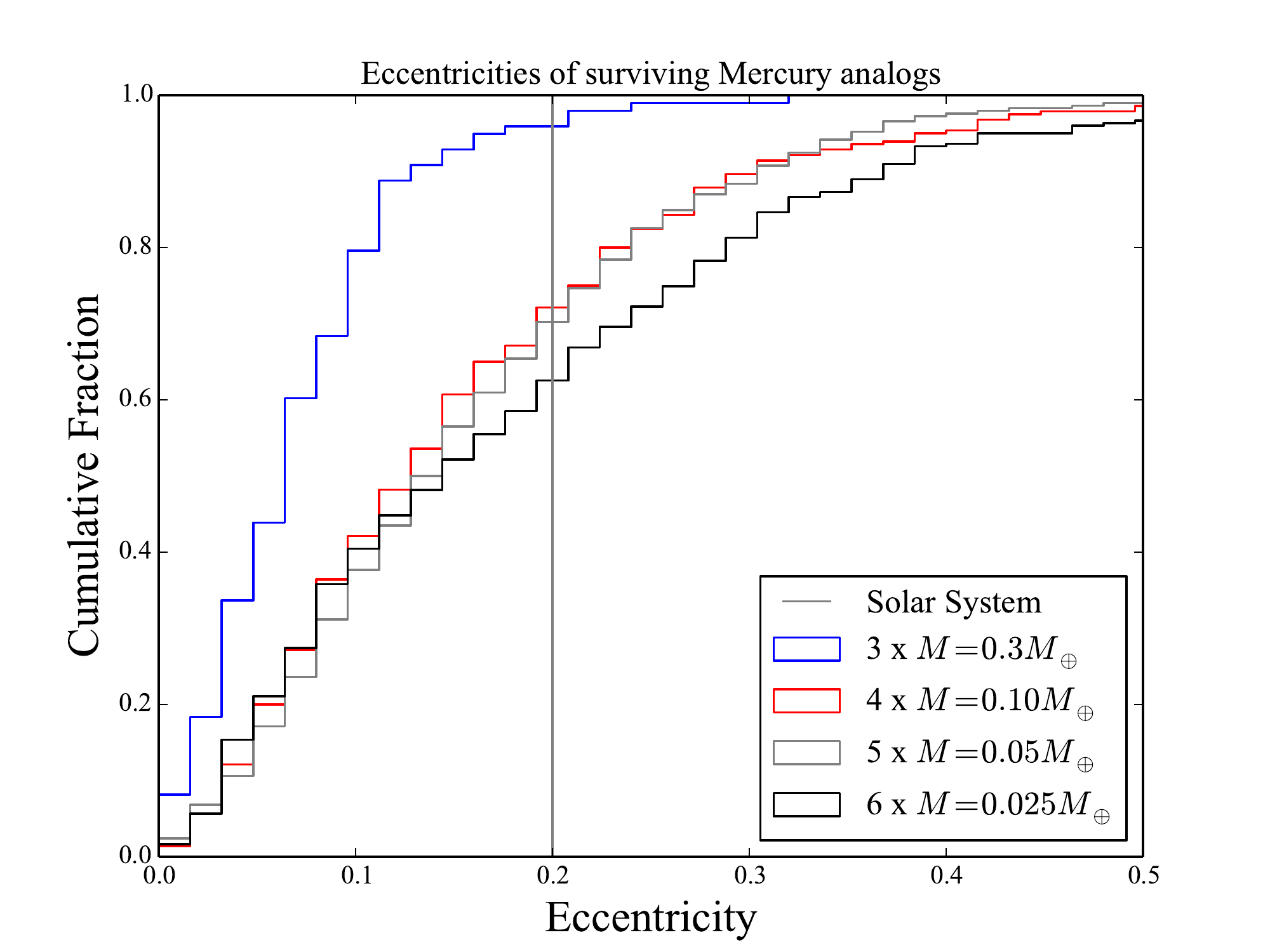}
	\caption{Cumulative fraction of final Mercury analog eccentricities in our various simulation sets.  The grey vertical line represents Mercury's modern eccentricity.}
	\label{fig:ecc}
\end{figure}

\begin{figure}
\centering
\includegraphics[width=.5\textwidth]{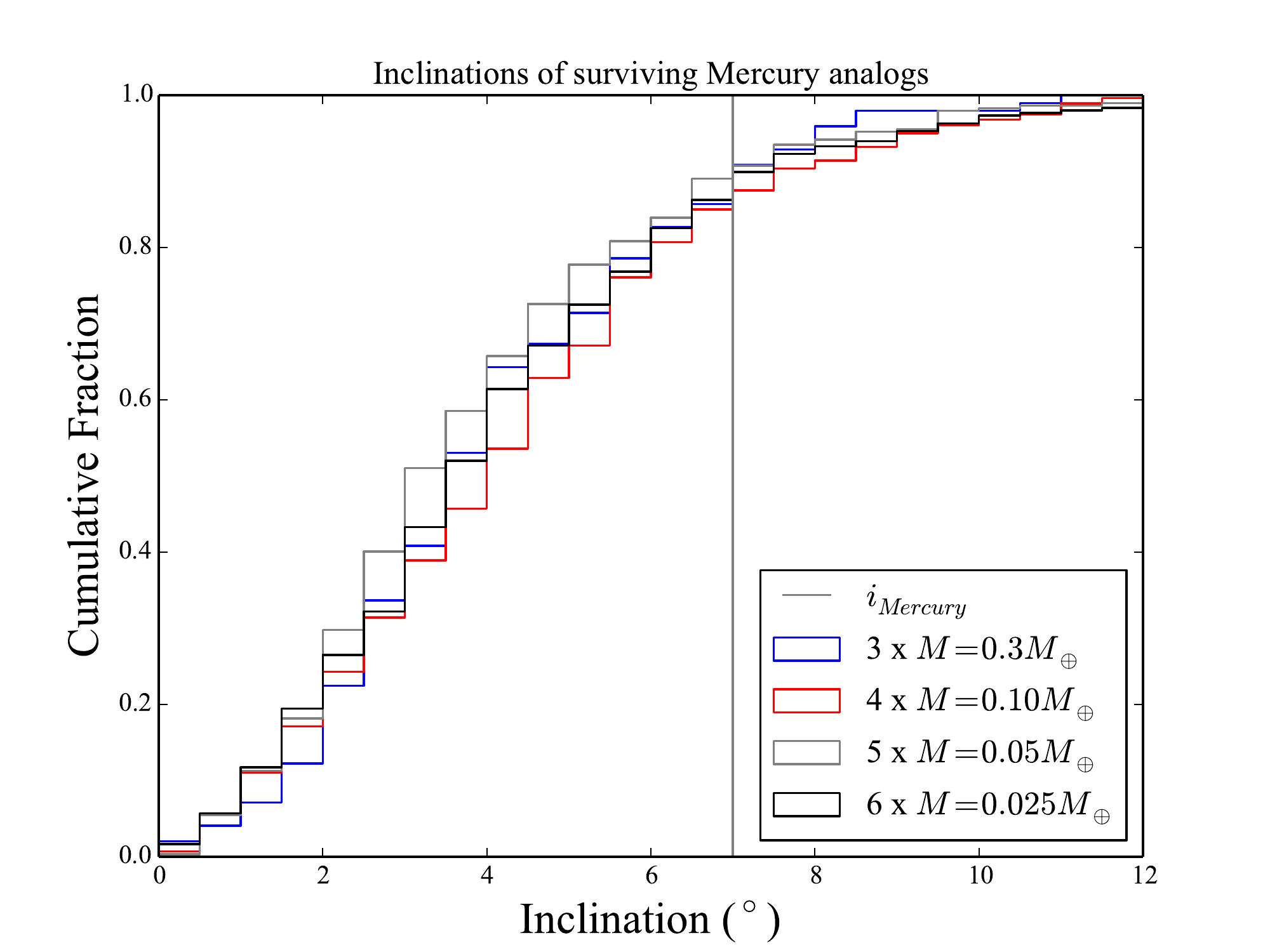}
	\caption{Cumulative fraction of final Mercury analog inclinations in our various simulation sets.  The grey vertical line represents Mercury's modern inclination.}
	\label{fig:inc}
\end{figure}

Figure \ref{fig:mercs} plots the final orbits of Mercury analogs in systems that finished with the correct number of planets (referred to throughout the subsequent text as 1-Mercury systems) for each of our different simulation batches.  It is clear that, regardless of initial conditions, Mercury's modern eccentricity and inclination are rather typical outcomes of our modeled scenario.  The cumulative distributions of eccentricities and inclinations plotted in figures \ref{fig:ecc} and \ref{fig:inc} demonstrate how our simulations do tend to slightly under-excite Mercury's orbit, though it is clear that our scenario yields a wide spectrum of outcomes.  It is still unclear whether Mercury's modern dynamical excitation is largely a consequence of the giant planet instability or the terrestrial planet formation process itself \citep[e.g.:][]{roig16,kaibcham16,clement18_frag}.  Our work presumes that the so-called Nice Model instability \citep{Tsi05,nesvorny12,deienno17} occurred prior to the loss of our hypothetical population of short-period planets \citep[consistent with an early instability as argued for by multiple recent studies:][]{morb18,clement18,nesvorny18,brasser20}.  While we leave the full study of our scenario's timing within the various proposed timelines for the terrestrial planets' early evolution to future work \citep[discussed further in section \ref{sect:discuss}; see recent reviews in:][]{izidoro18_book_review,ray18_rev}, we note that the Nice Model instability possibly played a role in sculpting Mercury's modern orbit.  Thus, while many of our final systems are remarkable solar system analogs, our scenario is by no means a comprehensive model for the early evolution of the inner solar system.

We compare the final mass and orbital period ratios of our Mercury-Venus analog systems to those of previous dynamical studies (figure \ref{fig:new_fig}) in figure \ref{fig:new_fig_tw}.  As in figure \ref{fig:mercs}, it is clear that the final semi-major axes of Mercury analogs in 1-Mercury systems decrease with decreasing initial proto-planet masses.  In the most extreme case ($M_{emb}=$ 0.025 $M_{\oplus}$) Mercury's modern semi-major axis lies outside of the range of values generated in our simulations.  In contrast, our higher-mass configurations of embryos (table \ref{table:ics}) are plagued by a low rate of instabilities and preponderance of overly massive Mercury analogs (e.g.: top left panel of figure \ref{fig:mercs}).  Thus, it is apparent that our configurations assuming more moderate initial embryo masses ($\sim$0.05-0.10 $M_{\oplus}$) are the most viable in terms of their ability to simultaneously match both Mercury's mass and orbit.  The magnitude of the eccentricity excitation induced on our proto-planets via secular resonant perturbations (section \ref{sect:destab}) from the other planets varies inversely with the mass of the primordial embryo \citep[e.g.:][]{dermott99}.  This is also evidenced by the fact that Mercury analogs produced in systems of less-massive proto-planets possess hotter $e$ and $i$ distributions (e.g.: figure \ref{fig:ecc}).  Thus, our $M_{emb}=$ 0.025 $M_{\oplus}$ embryos attain higher eccentricities during the instability, and are therefore more easily accreted by the other terrestrial planets.  Indeed, Venus and Earth accrete approximately three times as many embryos in our simulations investigating lower embryo masses than the higher-mass runs.  Thus, embryos in the lower-mass cases that begin interior to the $\nu_{5}$ resonance tend to merge and survive on an orbit that is closer to the Sun than in the actual solar system.  The embryos beginning these simulations exterior to $\nu_{5}$ are largely lost via collisions with Earth or Venus (discussed further in section \ref{sect:venus}).  The initial semi-major axes of the innermost 1-2 embryos also plays a minor role in producing the trend of final Mercury analogs surviving closer to the Sun in systems with less-massive proto-planets.  The inner planets in these systems originate further from $\nu_{5}$.  Thus, they are more likely to begin the integration with orbits that inhabit the more stable regions depicted in figure \ref{fig:map}.  However, on closer inspection we find that the innermost embryos never survive an instability without experiencing at least one collision with either a fragment or another embryo, thus we conclude that this is a minor effect.   

In contrast to our less massive systems, Mercury analogs produced from our $M_{emb}=$ 0.3 $M_{\oplus}$ runs inhabit a wider range of final semi-major axes.  Often in these cases either all three embryos survive, or the proto-planets combine with one another to form a single Mercury analog that is far too massive.  Finally, our $M_{emb}=$ 0.05-0.10 $M_{\oplus}$ sets occasionally finish with small planets exterior to $\nu_{5}$ ($a \gtrsim$0.35) like Mercury as the proto-planets are able to avoid collisions with Venus by attaining smaller eccentricities.  Moreover, the final analog planet masses in these simulations are closer to that of Mercury by virtue of the system beginning with less total mass in the region.

\begin{figure}
	\centering
	\includegraphics[width=.5\textwidth]{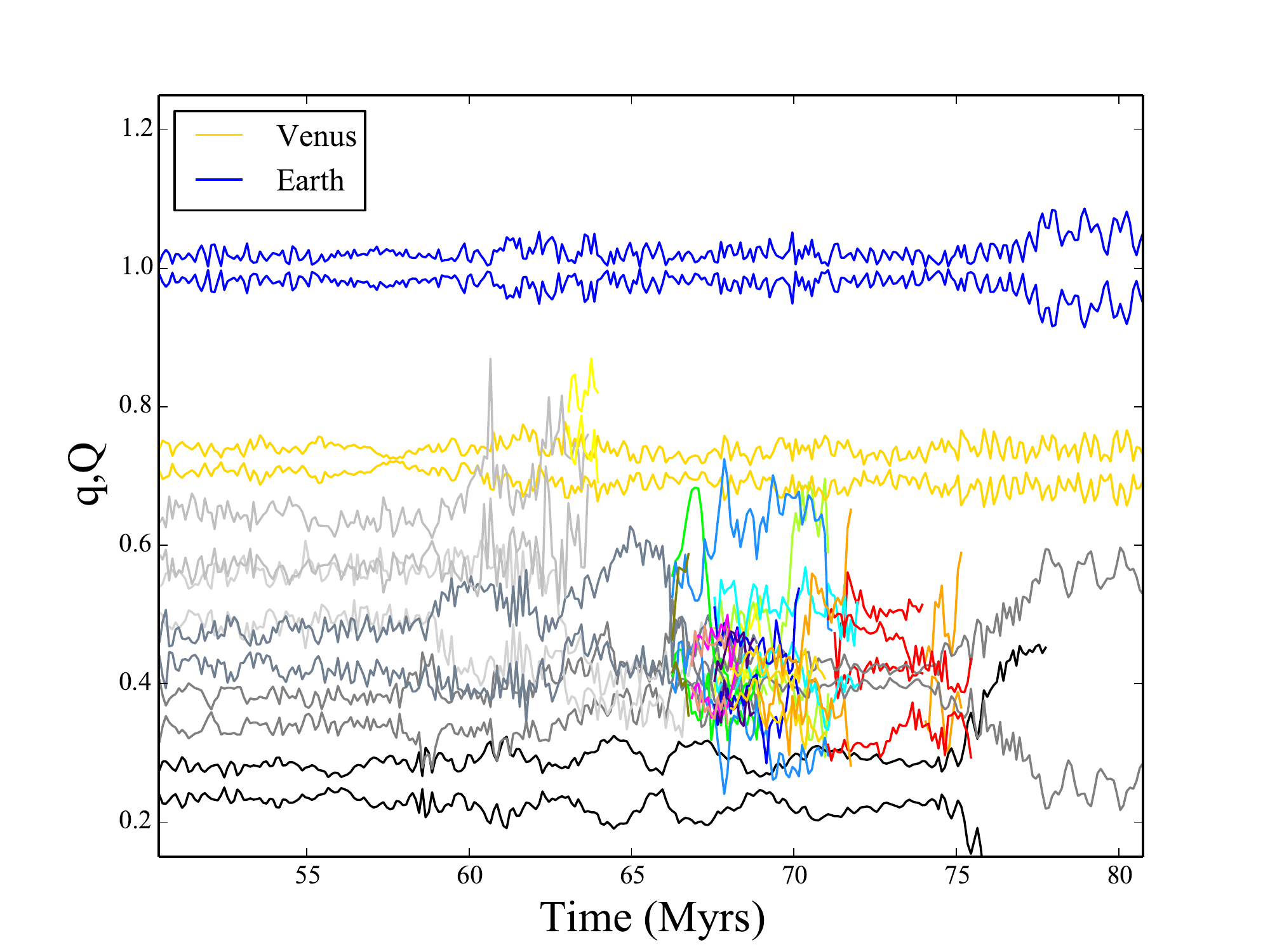}
	\caption{Example evolution of a system in the $M_{emb}=$ 0.05 $M_{\oplus}$ batch.  The perihelion and aphelion of each inner solar system object is plotted against simulation time.  The five initial embryos are plotted in different shades of grey, while each collisional fragment is assigned a random color.  The final Mercury analog has a mass of $M=$ 0.10 $M_{\oplus}$, a semi-major axis of 0.407 au ($a_{M,SS}=$ 0.387 au), eccentricity $e=$  0.33 ($e_{M,SS}=$ 0.21), inclination $i=$ 3.9$^{\circ}$ ($i_{M,SS}=$ 7.0$^{\circ}$), and CMF of 0.60.}
	\label{fig:qaq}
\end{figure}

An example evolution of a successful simulation from our $M_{emb}=$ 0.05 $M_{\oplus}$ batch is plotted in figure \ref{fig:qaq}.  The instability ensues at $t=$ 60 Myr when the second, third, and fourth embryos (in order of increasing distance from the Sun) undergo a series of six hit-and-run collisions.  Through this process, the most distant embryo is excited in eccentricity until it ultimately experiences a hit-and-run collision with Venus at $t=$ 63 Myr that produces 2 fragments.  All three remnant objects from this interaction are re-accreted by Venus within the subsequent $\sim$1.5 Myr, boosting the planets' mass by 0.05 $M_{\oplus}$.  3 more fragments are generated in a hit-and-run impact between the second and third embryo at $t=$ 66 Myr.  Swiftly following the injection of these additional particles, the system transforms rapidly through a series of $\sim$30 collisional interactions that produce an additional 12 fragments.  During this chaotic period of evolution, the embryo that goes on to become Mercury (second embryo) experiences a series of several imperfect accretion events that cumulatively boost its CMF.  In particular, a fragmenting collision between the ultimate Mercury-analog and the systems' fourth embryo (the mantle of which had already been significantly eroded in the $t=$ 66 Myr series of impacts) at $t=$ 68 Myr produces five new, mantle-only fragments.  Through this series of interactions, the CMF of the eventual Mercury analog is permanently altered to $\sim$0.60.  The mantle-rich fragments ultimately merge with the innermost embryo, which collides with the Sun at $t=$ 78 Myr.  It is also apparent from figure \ref{fig:qaq} that the orbits of Earth and Venus remain mostly quiescent during the instability.  We integrated the system for an additional 1 Gyr to confirm its dynamical stability.

\subsection{Consequences for Mercury}
\label{sect:merc}

\begin{figure*}
	\centering
	\includegraphics[width=.85\textwidth]{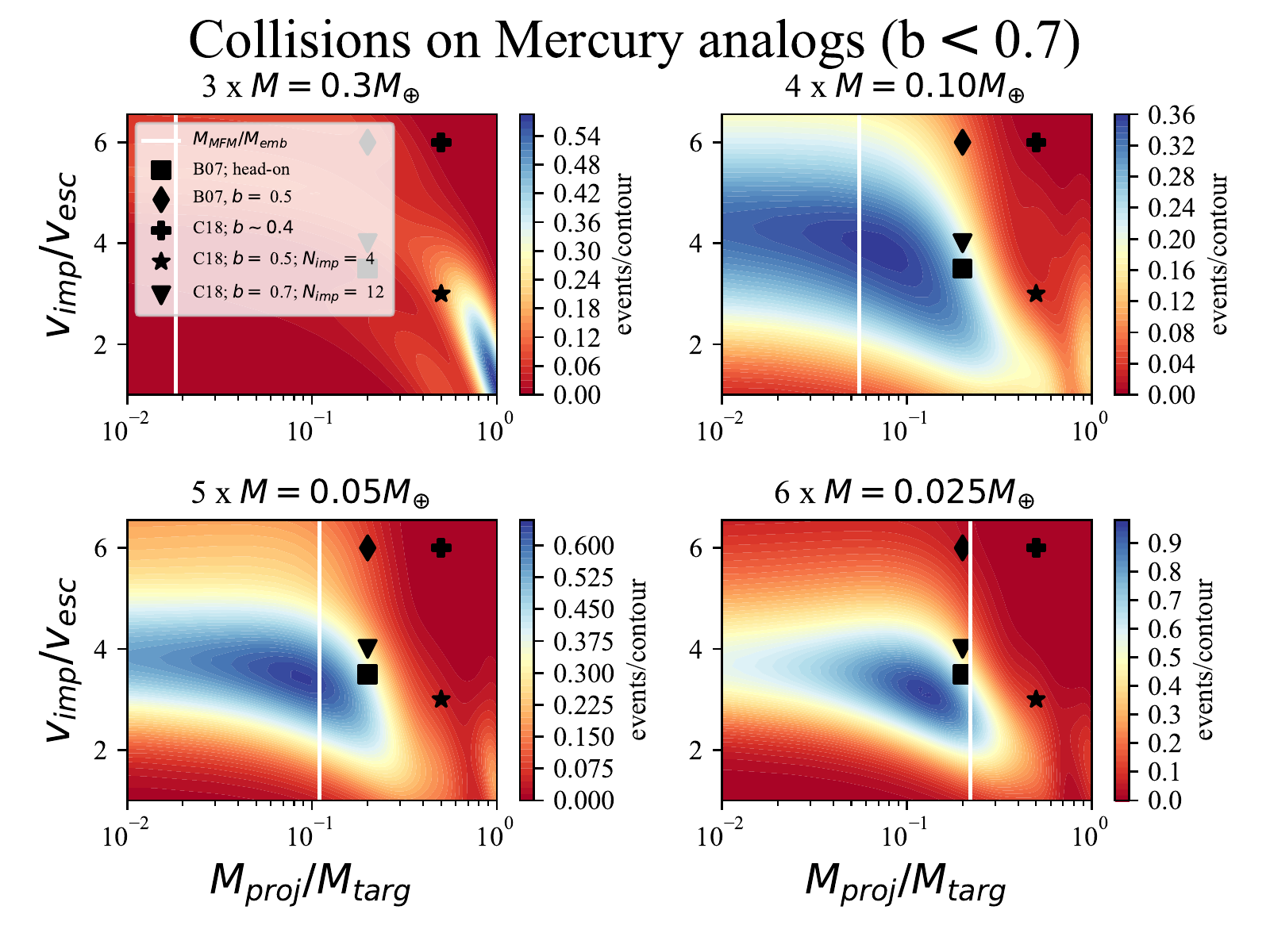}
	\caption{Relative frequency (kernel density estimate) of impact velocities (scaled by the mutual, two-body escape velocity) and projectile:target mass ratios in our various simulation sets.  The white vertical lines indicate the MFM:Embryo ($M_{MFM}/M_{emb}$) mass ratio in each simulation set.  The black symbols correspond to successful collisional geometries proposed in the literature.  Specifically, the square denotes a $b\simeq$ 0, head-on instance from \citet{benz07}, and the diamond indicates their preferred $b=$ 0.5 configuration.  The plus sign signifies the approximate preferred geometry for moderate $b$ in \citet{chau18}, while the star and upside down triangle provide two examples of multiple-impact scenarios from that work.}
	\label{fig:merc_col}
\end{figure*}

We now turn our attention to our Mercury analogs' collisional histories.  As our fragmentation algorithm \citep{chambers13} and corresponding CMF-tracking methodology (section \ref{sect:meth}) are based off the collisional regimes mapped in \citet{leinhardt12}, objects that emerge from our simulations with highly processed CMFs axiomatically experience one or more energetic, mantle-stripping collisional encounters \citep{benz88,jackson18,chau18}.  Nevertheless, it is worthwhile to discuss the collisional environment within our computations in the context of hydrodynamical studies specifically devised to transform proto-Mercury's bulk density.  As discussed in the introduction, one possibility is that Mercury was the projectile striking a much larger body \citep[either proto-Venus, proto-Earth, or an object that no longer resides in the solar system:][]{asphaug14}.  We analyzed the dynamical problems with this scenario in detail in \citet{clement19_merc}.  A second plausible circumstance is one where proto-Mercury (near its current orbit) is struck by a smaller body.  \citet{benz07} analyzed this possibility and found that the collisional velocities required to erode the mantle of a 0.12 $M_{\oplus}$ (2.25 times Mercury's modern mass) version of Mercury are typically high ($\gtrsim$6$v_{esc}$; diamond in figure \ref{fig:merc_col}).  However, the authors obtained one moderately successful case from a nearly head-on impact at $\sim$4$v_{esc}$.  More recently, \citet{chau18} thoroughly investigated the applicable collisional parameter space in the classic \citet{benz07} scenario, and favored a lower-velocity ($\sim$4$v_{esc}$) collision and less-extreme impact geometry (impact parameter, $b\simeq$ 0.4).  Additionally, the authors explored the possibility that Mercury's mantle was eroded in a series of similar collisions, and tabulated the number of impacts required for given combinations of $M_{proj}/M_{targ}$, $b$ and $v_{imp}/v_{esc}$.  Figure \ref{fig:merc_col} plots the relative frequency of impact velocities and $M_{proj}/M_{targ}$ for all collisions on proto-planets in our various simulation batches, compared with a range of successful scenarios from \citet{benz07} and \citet{chau18}.  While the typical target particle masses in our simulations investigating $M_{emb}=$ 0.3 and 0.025 $M_{\oplus}$ are different than what is invoked in the classic single-erosive impact model (a fully differentiated body with $\sim$2.25 times Mercury's modern mass), those in our more dynamically successful batches testing $M_{emb}=$ 0.1 and 0.055 $M_{\oplus}$ are quite similar.  Interestingly, the most common combinations of $M_{proj}/M_{targ}$ and $v_{imp}/v_{esc}$ in these sets (blue regions in figure \ref{fig:merc_col}) necessitate a head-on collision (black square) to adequately reshape Mercury's mantle.  However, a multiple-impact scenario with a less-extreme orientation (upside-down triangle) is certainly plausible in our model.  In such a scenario, Mercury would experience multiple erosive encounters with other proto-planets and collisional debris in the region.

\begin{figure}
	\centering
	\includegraphics[width=.5\textwidth]{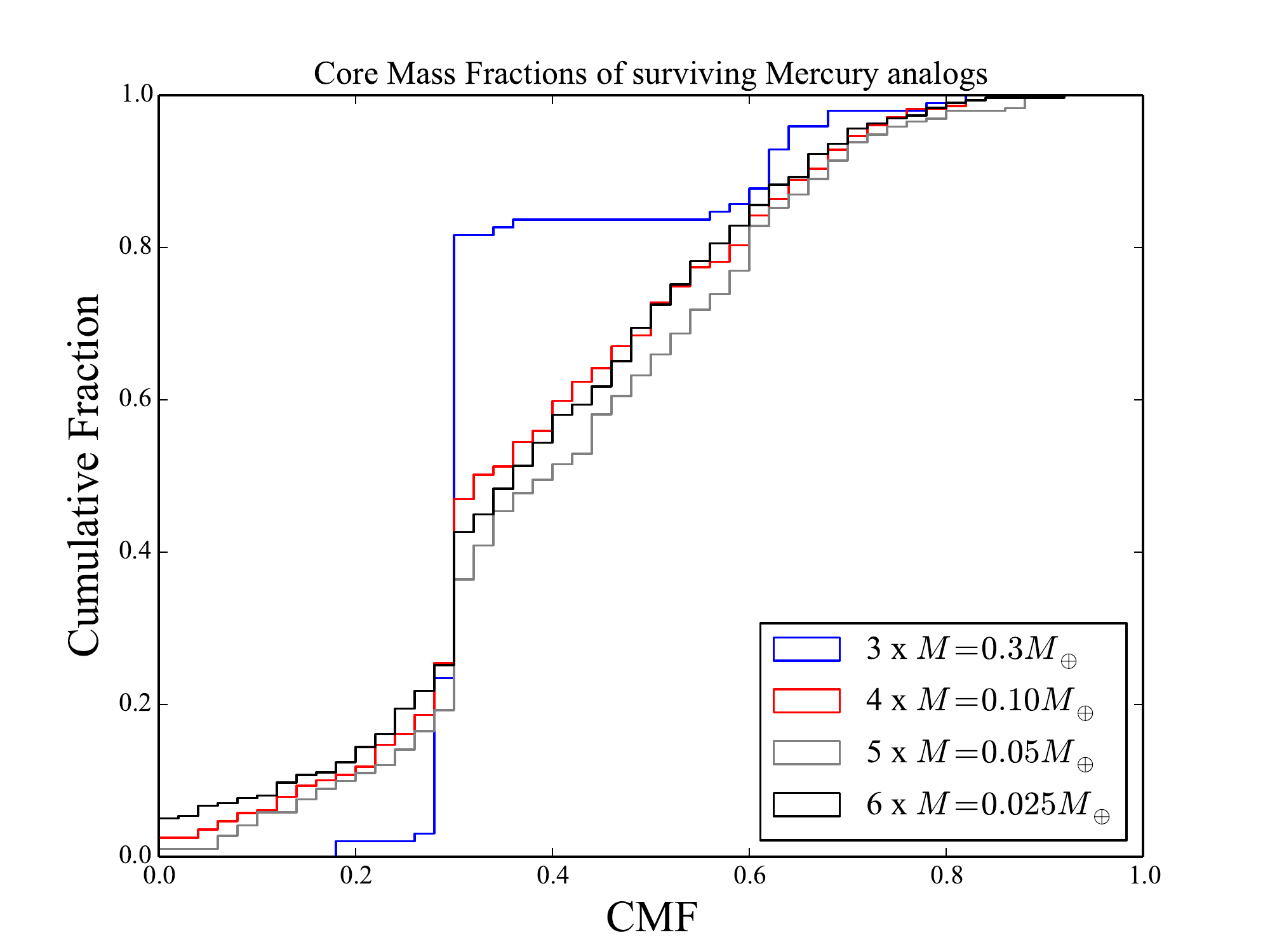}
	\caption{Cumulative distribution of Mercury analog CMFs in our various simulation sets.}
	\label{fig:cmf}
\end{figure}

To evaluate our scenario's success in terms of its ability to collisionally strip Mercury's mantle, we plot the final CMFs of all surviving Mercury analogs in our simulations in figure \ref{fig:cmf}.  It is clear that, with the exception of our $M_{emb}=$ 0.3 $M_{\oplus}$ set where instabilities occur quite infrequently, our systems' populations of Mercury analogs are highly processed and altered in CMF via the collisional interactions that occur during the instability.  Indeed, only $\sim$20$\%$ of surviving Mercury analogs in our three sets testing lower values of $M_{emb}$ retain their original CMFs.  Thus, the 1-Mercury systems depicted in figure \ref{fig:mercs} are quite diverse in terms of Mercury's CMF.  However, we remind the reader that these results are partially a consequence of our code's fragmentation algorithm (section \ref{sect:meth}).  In particular, when a fragmenting collision occurs between two lower-mass embryos, the resulting collisional fragments have masses closer to those of the embryos themselves as we utilize the same MFM (0.0055 $M_{\oplus}$) in all of our simulations.  Follow-on collisions in these systems can still be erosive and thus still CMF-altering over a wider range of collisional velocities \citep[e.g.:][]{leinhardt12,genda12,gabriel20}.

\begin{figure}
    \centering
	\includegraphics[width=.5\textwidth]{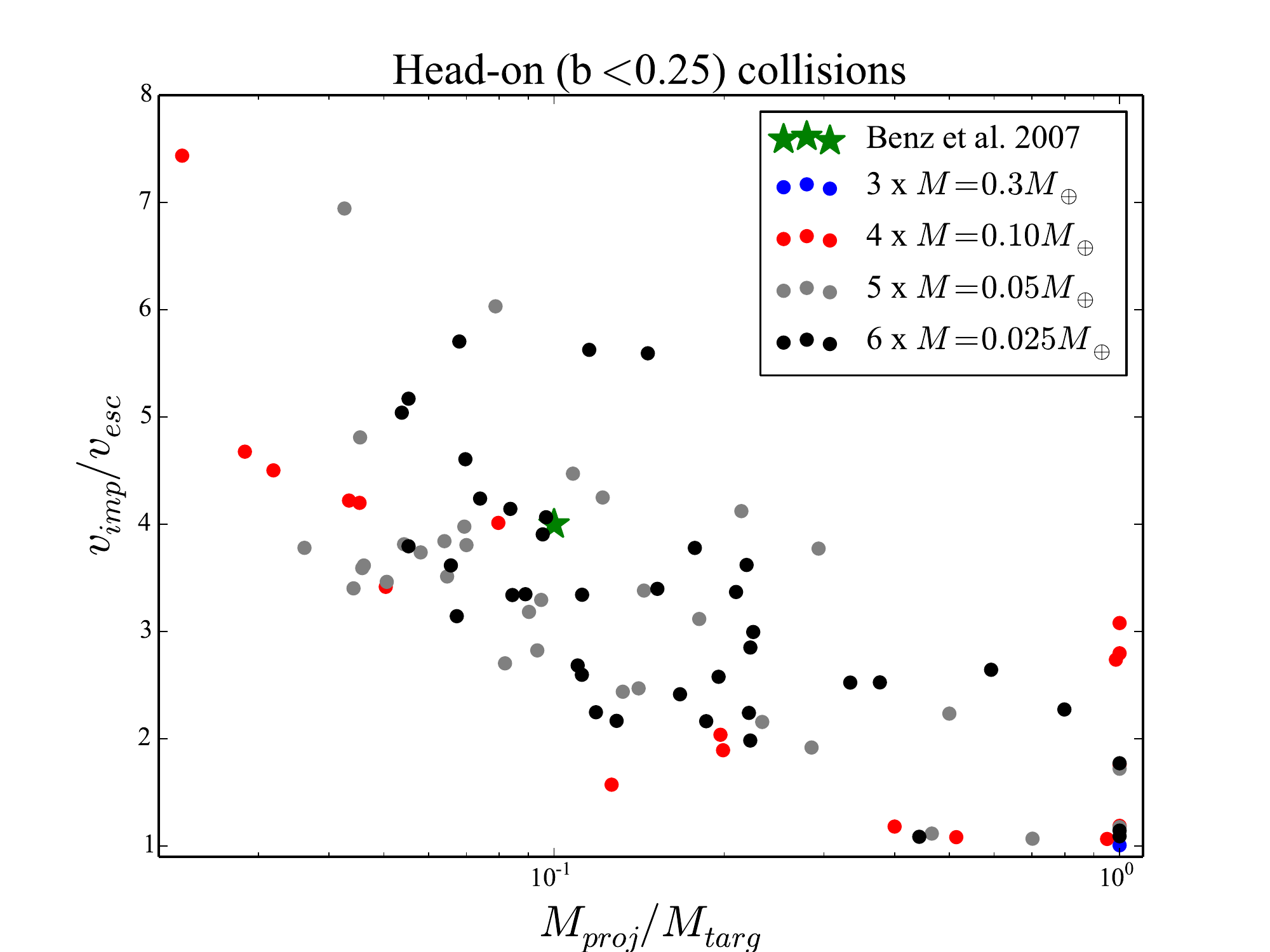}
	\caption{Head-on collisions in our simulations (here defined as impact parameters less than 0.25) similar to that of the model of \citet{benz07} where Mercury's mantle is eroded by a single impact of a smaller projectile.  The color of each point corresponds to the simulation set (table \ref{table:ics}) in which the impact occurred.  The successful head-on case from \citet{benz07} (run 6 in that work) is plotted with a green square.}
	\label{fig:merc_col2}
\end{figure}

At this point, it is reasonable to question whether our fragmentation model provides an accurate representation of the problem we seek to investigate.  Hydrodynamical models tend to find that the majority of the disrupted material in giant impacts similar to those studied in this manuscript should be ejected in the form of dust.  Indeed, observed debris disks \citep[e.g.:][]{lisse09,weinberger11} interpreted to be potential tracers of the giant impact process of terrestrial planet formation \citep{genda15} are largely consistent with this assumption.  Thus, while a prescription for tracking fragmentation events within our N-body code \citep{leinhardt12,chambers13} is undoubtedly necessary for modeling mantle loss on our Mercury analogs, it is at best unclear whether tracking the ejecta as massive ($\sim$Lunar-mass), dynamically interacting bodies is physical.  However, we note that, in all four of our simulation batches, $\gtrsim$90$\%$ of produced fragments are subsequently removed via a perfectly accretionary collision without ever undergoing subsequent fragmentation events.  In this manner, the vast majority of our high-CMF Mercury analogs acquire their large Fe/Si ratios through a small-number of fragmenting collisions that tend to occur with low impact parameters \citep[similar to the head-on scenario of ][see figure \ref{fig:merc_col2}]{benz07}.  The mantle-rich fragments produced in these impacts dynamically interact with the other objects in the simulation until they are eventually accreted by \textit{another planet}.  If this ejected material were in the form of dust rather than massive bodies, it is reasonable to argue that it too might be carried away from an eventual Mercury analog via PR drag \citep{melis12} or interactions with the solar wind \citep{spalding20} rather than being re-accreted \citep[e.g.:][]{gladman09}.  If this were the case, our simulations should systematically \textit{overestimate} both the masses and \textit{underestimate} the CMFs of Mercury analogs (as disrupted material would be re-accreted less frequently).  Thus, we contend that our CMF results are not an artifact of our computation methodology as repeated collisions between fragments do not appreciably contribute to the final CMFs of our Mercury analogs.

While our simulations are quite successful at generating a diverse range of Mercury analog Fe/Si compositions (figures \ref{fig:mercs} and \ref{fig:cmf}), as discussed in section \ref{sect:analog}, the surviving analogs themselves tend to be positioned on the sunward side of the $\nu_{5}$ resonance.  However, our integrations begin with the giant planets on their modern orbits \citep[i.e., after the giant planet instability:][]{Tsi05,nesvorny12}.  Thus, if our formation scenario occurred either prior to, or in conjunction with the Nice Model instability \citep[e.g.:][]{clement18}, the $\nu_{5}$ resonance would be located further from the Sun by virtue of Jupiter's increase rate of perihelia precession \citep[induced by stronger mutual interactions with Saturn:][]{clement20_mnras}.  This might boost the likelihood of Mercury's survival near its modern semi major axis (0.387 au) exterior to $\nu_{5}$, provided the diminutive planet could withstand the dynamical perturbations involved with $\nu_{5}$ migrating through its orbit during the instability \citep{roig16}.  In this same manner, if our scenario pre-dated the Moon-forming impact \citep[e.g.:][]{earth,kleine09,canup12} the resonant perturbations in the $a<$ 0.6 au region induced by the terrestrial planets would obviously be different than those modeled in our simulations by virtue of the system possessing additional large terrestrial bodies.  Thus, a complete validation of our hypothetical evolutionary sequence must study its viability throughout the various stages of solar system evolution, and particularly focus on the consequences of the giant planet instability.

\subsection{Consequences for Venus}
\label{sect:venus}

\begin{figure}
	\centering
	\includegraphics[width=.5\textwidth]{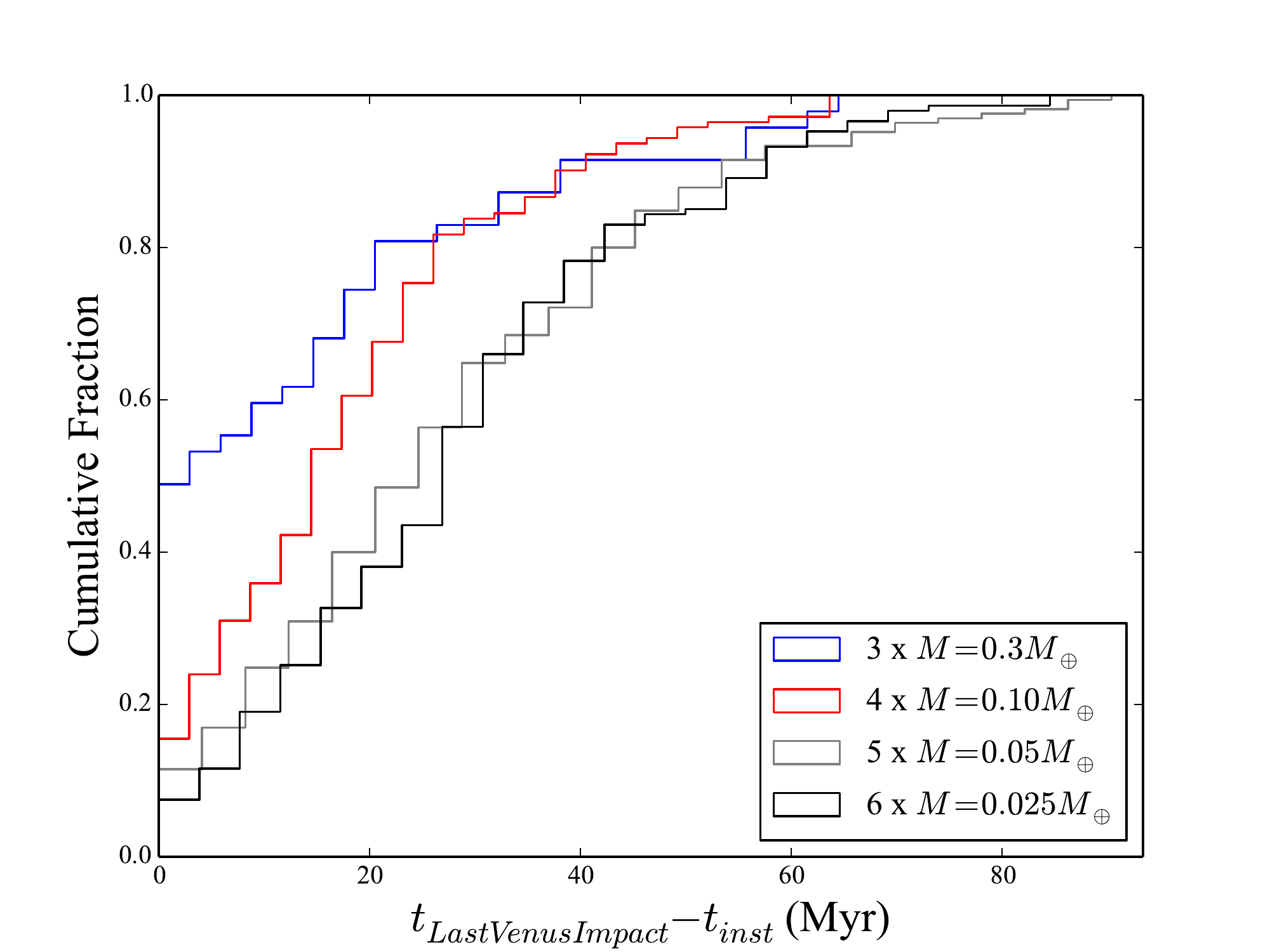}
	\caption{Cumulative distribution of final impact times (relative to the instability time: figure \ref{fig:tinst}) on Venus in our various simulation sets.}
	\label{fig:tinst_ven}
\end{figure}

A significant consequence of our scenario is a delayed accretion event on Venus.  While not ubiquitous, these delayed mass additions occur on timescales of some 20-30 Myr (figure \ref{fig:tinst_ven}) after the onset of the instability (figure \ref{fig:tinst}).  However, as our simulation batches investigating lower values of $M_{emb}$ litter the inner solar system with significantly more debris in the form of fragments generated in repeated imperfect collisions (section \ref{sect:merc}), the ultimate merger with Venus can occur $\gtrsim$60-80 Myr after the onset of the instability.  In general, the total mass of additional material accreted by Venus (and, to a lesser extent, Earth) is proportional to the total initial mass of the system's proto-planets.  Venus gains an average of 0.13 $M_{\oplus}$ of material after the instability in our batch testing a configuration of three, $M_{emb}=$ 0.3 $M_{\oplus}$ objects.  Conversely, the planet is only impacted by an average of 0.043, 0.022, and 0.017 $M_{\oplus}$ worth of embryos and debris fragments in our systems testing proto-planet masses of 0.1, 0.05 and 0.025 $M_{\oplus}$, respectively.  In addition to these delayed accretion events, Venus also occasionally experiences a small number of hit-and-run collision ($\sim$35-42$\%$ of all collisions on Venus, depending on the simulation batch) en route to eventually accreting the impacting material.  While Venus' ultimate structure would be rather unaffected by this additional delivery of mass (presuming a chondritic composition)  if our scenario occurred early in the solar system's history \citep[in conjunction with terrestrial planet formation:][]{clement18}, the fact that Venus does not possess an internally generated magnetic field or natural satellite is potentially at odds with primordial stratification in its core having been disrupted by a late giant impact as was the case on Earth \citep{jacobson17b}.  On closer inspection, it is clear that the vast majority ($\gtrsim$90$\%$) of the material impacting Venus in our simulations begins the simulation outside of Mercury's modern orbit.  Thus, it might be easy to prevent late impacts on Venus by altering the range of initial semi-major axes for our proto-planets (i.e., not placing any embryos at $a\gtrsim$0.5 au).  However, such a mass configuration is quite ad hoc, and it is unclear how such a wide gap in the surface density profile of our proto-planets would have developed.  Thus, we favor a uniform distribution of embryos as our modeled evolutions provide a compelling explanation for the modern Mercury-Venus period ratio.

\begin{figure}
	\centering
	\includegraphics[width=.5\textwidth]{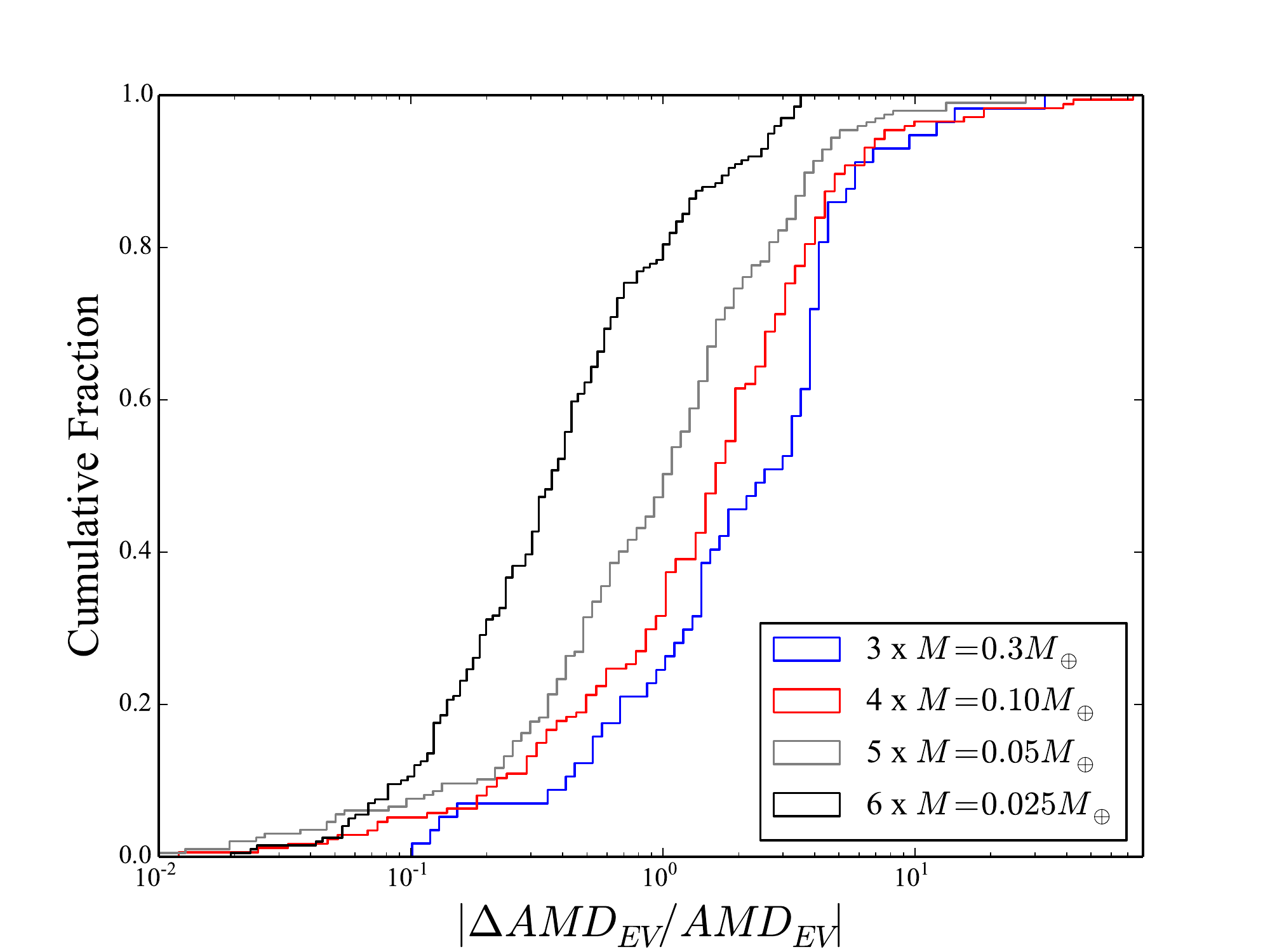}
	\caption{Cumulative distribution of the fractional change in Venus and Earth's AMD after the instability in our various simulation sets.}
	\label{fig:amd}
\end{figure}

While Venus' mass is somewhat altered in our scenario, the effect on the other terrestrial planets' orbits is rather minimal, though not negligible.  Figure \ref{fig:amd} plots the fractional change in angular momentum deficit (AMD) for Earth and Venus in our various simulation sets:

\begin{equation}
	AMD = \frac{\sum_{i}M_{i}\sqrt{a_{i}}[1 - \sqrt{(1 - e_{i}^2)}\cos{i_{i}}]} {\sum_{i}M_{i}\sqrt{a_{i}}} 
	\label{eqn:amd}
\end{equation}

The majority of the terrestrial planet AMDs in our simulations testing higher values of $M_{emb}$ are altered (almost exclusively increased) by over a factor of two.  In some $\sim$5$\%$ of cases, $AMD_{EV}$ is increased by a whole order of magnitude.  Thus, our results indicate that larger initial values of $M_{emb}$ are potentially incompatible with the modern, dynamically cold orbits of Earth and Venus.  While a small fraction of systems in all four of our simulation batches do experience only minor changes in $AMD_{EV}$, a majority of the configurations considering smaller values of $M_{emb}$ undergo only mild alterations in the orbits of Earth and Venus.  Thus, our results indicate that a version of the \citet{volk15} scenario incorporating smaller proto-planets is perhaps more promising given the improved likelihood of systems experiencing smaller net changes in $AMD_{EV}$, and the systematically lower masses of Mercury analogs in systems finishing with the correct number of planets (figure \ref{fig:mercs}).

\subsection{Proto-Mercury stability during terrestrial planet formation}

It is worthwhile to investigate our systems of proto-planets' stability during the chaotic giant impact epoch of terrestrial planet formation, and similarly understand whether it is reasonable to disentangle our proposed scenario from Earth and Venus' formation as assumed in section \ref{sect:analog}.  To accomplish this, we perform an additional suite of 50 simulations where we place our systems of proto-planets interior to a terrestrial forming disk of embryos and planetesimals.  For simplicity, we distribute the planetesimals and embryos in a narrow annulus between 0.75-1.05 au \citep[e.g.:][]{hansen09,walsh16,lykawka20}.  While the process of terrestrial planet formation is rather complex, and a subject of ongoing research, we select these initial conditions to maximize the possibility of forming Venus, Earth and Mars-like planets at the proper radial locations, and minimize the computational cost of the calculation \citep[for a recent review see:][]{ray18_rev}.  Each system considers 2.0 $M_{\oplus}$ of planet-forming material; distributed such that half the mass is in 12 equal-mass embryos and the remaining half is in 200 equal-mass planetesimals.  Orbits for these particles are selected using the methodology described in \citet{clement18_frag}.  In 25 simulations, we study the evolution of our 4 x 0.1 $M_{\oplus}$ proto-planet systems, and in an additional 25 runs we follow the dynamics of our 5 x 0.05 $M_{\oplus}$ proto-planet initial conditions (table \ref{table:ics}).  Each system is integrated for 100 Myr using the same Mercury6 integrator and timestep employed throughout our manuscript \citep{chambers99}.  However, these additional simulations do not incorporate a fragmentation algorithm.

\begin{figure}
	\centering
	\includegraphics[width=.5\textwidth]{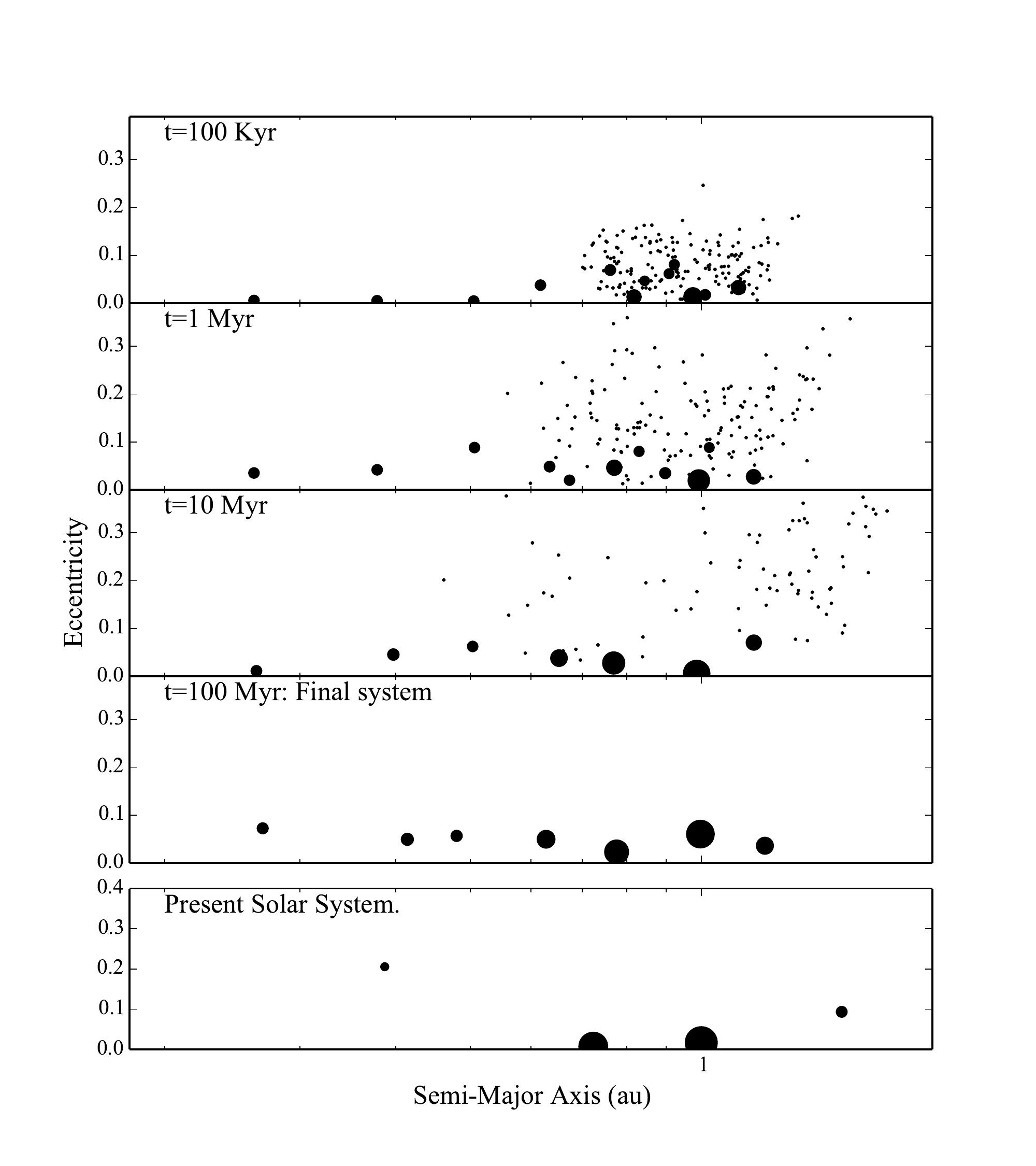}
	\caption{Semi-Major Axis/Eccentricity plot depicting the evolution of a simulation where a system of 4 $M_{emb}=$ 0.1 $M_{\oplus}$ proto-planets remain stable for 100 Myr during the giant impact phase of terrestrial planet formation. The size of each point corresponds to the mass of the particle. The final planet masses are 0.57, 0.76 and 0.28 $M_{\oplus}$, respectively.  The three inner proto-planets accrete a small number of planetesimals during the simulation, altering their masses to 0.115, 0.115 and 0.135 $M_{\oplus}$, respectively.  The outermost proto-planet accretes several planetesimals and two embryos from the Venus-forming region, thereby increasing its' mass to 0.313 $M_{\oplus}$.}
	\label{fig:tp_evo}
\end{figure}

In general our additional simulations evince a broad spectrum of outcomes.  In fact, the final states of four of our additional runs are qualitatively similar to those attained by our embryos integrated within the modern solar system (section \ref{sect:analog}).  Thus, the outer 1-2 proto-planets incorporate into Venus, and the inner 2-3 merge in a series of high-energy collisions to form a dynamically isolated Mercury analog.  More typically, however, the outer 1-2 embryos merge with the forming Venus, and the innermost proto-planets survive the 100 Myr integration on relatively unaltered orbits.  Indeed, the median number of surviving proto-planets among our 50 terrestrial planet formation simulations is 3.  While we observe a few cases where the outermost proto-planet does not merge with Venus, and instead combines with one of the interior proto-planets, the overwhelming number of systems largely retain their initial proto-planet structures with the exception of objects that merge with Venus.  Indeed, one proto-Mercury system of four 0.1 $M_{\oplus}$ embryos survives the entire process of terrestrial planet formation intact (aside from each proto-planet accreting material scattered out of the terrestrial-forming disk).  The evolution of this successful simulation is plotted in figure \ref{fig:tp_evo}.

It is far beyond the scope of this paper to comprehensively study the survivability of our Mercury-forming proto-planets within the various proposed terrestrial formation \citep[e.g.:][]{walsh11,bromley17,ray17sci,clement18} and giant planet migration sequences \citep[e.g.:][]{nesvorny12,pierens14,ray17sci,ribeiro20,clement21_instb}.  Nevertheless, we present these initial simulations as a proof-of-concept in order to demonstrate that the dynamical evolution of the embryos in our proposed scenario (particularly the innermost ones that tend to form the final Mercury analog) is not strongly coupled to the process of terrestrial planet formation.

\section{Discussion}
\label{sect:discuss}

While our proposed scenario is quite successful at consistently generating instabilities (figure \ref{fig:tinst}) and energetic collisions between proto-planets that often produce high-CMF objects that ultimately survive as Mercury analogs (figure \ref{fig:mercs} and \ref{fig:cmf}), we acknowledge that our initial conditions are highly idealized.  In particular, we assume that the giant planets already acquired their modern orbits through the Nice Model instability before the onset of our proposed scenario.  However, we argue that the important conclusion of our work is that large swaths of orbital parameter space interior to Venus are rather unstable due to a high density of overlapping secular and mean motion resonances.  If additional, short-period planets had formed in this region, our simulations indicate that instabilities among the hypothetical bodies are fairly ubiquitous (even in absence of an external trigger, discussed further below).  Such a scenario is particularly compelling because it provides a natural explanation for Mercury's high mean density and dynamical offset from Venus.  Moreover, if the initial planets' masses are small enough, the resulting instabilities leave only a small dynamical footprint on Earth and Venus.  Thus, our scenario would presumably still be viable if it unfolded while the other terrestrial planets were still forming.

We can now speculate that a comprehensive model for the formation of the inner solar system might be one where a single dynamical instability sculpts the entire solar system in a single, decisive event.  Indeed, it is quite cumbersome to invoke multiple epochs of quasi-stability and more than one episode of dynamical instability in different regions of the solar system over various time periods.  In this manner, a Nice Model instability that ensues rather swiftly following nebular gas dispersal as the result of the giant planets having formed in an unstable configuration \citep{ribeiro20,clement21_instb} is philosophically appealing.  While certain constraints \citep[for example, the Kuiper Belt's orbital distribution:][]{nesvorny15a,nesvorny15b} speak against such an instability occurring within the first few Myr after gas disk dissipation, such a series of events is capable of disrupting embryos in the Mars-forming region in a manner such that the planet's final mass is well-reproduced \citep{ray09a,clement18,clement18_frag,nesvorny21}.  Thus, one might imagine a configuration of quasi-stable 0.1-0.3 $M_{\oplus}$ embryos \citep{walsh19,clement20_psj,woo21} in the inner solar system (extending inward to $\sim$0.2 au as proposed in this paper) emerging from the gas disk being perturbed by the giant planet instability; thereby triggering the final giant impacts on Earth and Venus \citep[and eventually leading to the Moon's formation:][]{canup12}, limiting Mars' growth \citep{Dauphas11,clement18,nesvorny21}, depleting and exciting the asteroid belt \citep{deienno18,clement18_ab}, and leaving Mercury as the sole survivor of the primordial generation of short-period embryos \citep{volk15}.  While a robust study of such a complex dynamical scenario is beyond the scope of our present manuscript, we stress that it is vitally important for any model of planet formation in the solar system to simultaneously reconcile Mercury's peculiar size, composition and orbit.

\section{Conclusions}

We presented an analysis of a hypothetical scenario where Mercury emerges as the lone survivor of a dynamical instability among a primordial generation of short-period proto-planets in the inner solar system.  We find that such a series of events is compelling in terms of its ability to consistently alter Mercury's Fe/Si ratio and produce a large dynamical offset between Venus and the surviving Mercury analog \citep[often in excess of the modern value of $P_{V}/P_{M}=$ 2.6, a significant improvement from previous models:][]{clement19_merc}.  Additionally, our simulations indicate that instabilities between proto-planets with masses in excess of a few times that of Mars tend to over-excite the orbits of Earth and Venus, and consistently yield Mercury analogs that are an order of magnitude too massive.  These systems also deliver a significant amount material to Venus after the instability, which may be problematic if primordial stratification in Venus' core is the reason for its lack of an internally generated magnetic dynamo \citep{jacobson17b}.  In contrast, our integrations considering proto-planet masses of $\sim$0.025-0.1 $M_{\oplus}$ are more promising in terms of their success when measured against both of the aforementioned constraints.  However, the surviving Mercury-analogs in these more diminutive architectures almost exclusively inhabit semi-major axes interior to the modern location of the $\nu_{5}$ secular resonance (inside of Mercury's actual orbit).  We argue this effect might not be as severe if the system of short-period proto-planets destabilized before or during the giant planet instability as the $\nu_{5}$ resonance was oriented differently in the solar system's infancy.  While our work indicates that the destruction of a primordial generation of short-period proto-planets is a potentially viable avenue for Mercury's origin, substantial additional modeling work will be required to comprehensively validate its feasibility within the larger context of terrestrial planet formation.

\section*{Acknowledgments}

We are thank Nathan Kaib, Sean Raymond and an anonymous reviewer for useful and insightful discussions.  We are also grateful to Andr\'{e} Izidoro, Seth Jacobson and Patryk Lykawka for graciously and promptly sharing data their from \citet{izidoro15}, \citet{jacobson14} and \citet{lykawka19} for the production of figures \ref{fig:new_fig} and \ref{fig:new_fig_tw}.  The majority of computing for this project was performed at the OU Supercomputing Center for Education and Research (OSCER) at the University of Oklahoma (OU).  This research was done using resources provided by the Open Science Grid \citep{osg1,osg2}, which is supported by the National Science Foundation award 1148698, and the U.S. Department of Energy's Office of Science.

\bibliographystyle{apj}
\newcommand{\sci}{$Science$ }
\bibliography{merc3}
\end{document}